\documentclass[apj]{emulateapj}

\newcommand{\etal}{{et al.~}}
\newcommand{\msunh}{\>h^{-1}\rm M_\odot}
\newcommand{\Msun}{\>{\rm M_\odot}}

\newcommand{\calC}{{\cal C}}
\newcommand{\rmag}{\>^{0.1}{\rm M}_r-5\log h}

\shorttitle{Galaxy Groups in SDSS DR4: II}

\shortauthors{Yang et al.}

\begin{document}


\title{Galaxy Groups in the SDSS DR4: II. halo occupation statistics}

\author{Xiaohu Yang\altaffilmark{1,4}, H.J. Mo \altaffilmark{2},
        Frank C. van den Bosch\altaffilmark{3}}

      \altaffiltext{1}{Shanghai Astronomical Observatory,
         the Partner Group of MPA, Nandan Road 80, Shanghai 200030, China;
         E-mail: xhyang@shao.ac.cn}
      \altaffiltext{2}{Department of Astronomy, University of Massachusetts,
        Amherst MA 01003-9305}
      \altaffiltext{3} {Max-Planck-Institute for Astronomy, K\"onigstuhl 17,
        D-69117 Heidelberg, Germany }
      \altaffiltext{4}{Joint Institute for Galaxy and Cosmology (JOINGC) of
        Shanghai Astronomical Observatory and University of Science and
        Technology of China}


\begin{abstract} We  investigate various galaxy occupation  statistics of dark
matter halos using  a large galaxy group catalogue  constructed from the Sloan
Digital Sky Survey Data Release 4 (SDSS DR4) with an adaptive halo-based group
finder.   The  conditional  luminosity  function (CLF),  which  describes  the
luminosity  distribution of galaxies  in halos  of a  given mass,  is measured
separately for all, red and blue galaxies,  as well as in terms of central and
satellite galaxies.  The  CLFs for central and satellite  galaxies can be well
modelled  with  a  log-normal  distribution  and a  modified  Schechter  form,
respectively.   About 85\%  of  the central  galaxies  and about  80\% of  the
satellite  galaxies  in  halos  with  masses $M_h\ga  10^{14}\msunh$  are  red
galaxies. These numbers decrease to 50\% and 40\%, respectively, in halos with
$M_h \sim 10^{12}\msunh$.  For halos of  a given mass, the distribution of the
luminosities of central galaxies, $L_c$, has a dispersion of about $0.15$ dex.
The mean  luminosity (stellar mass) of  the central galaxies  scales with halo
mass as $L_c \propto M_h^{0.17}$ ($M_{*,c} \propto M_h^{0.22}$) for halos with
masses $M\gg  10^{12.5}\msunh$, and  both relations are  significantly steeper
for less  massive halos.   We also measure  the luminosity (stellar  mass) gap
between the first  and second brightest (most massive)  member galaxies, $\log
L_1  -  \log  L_2$  ($\log  M_{*,1}-\log  M_{*,2}$).   These  gap  statistics,
especially  in  halos with  $M_h  \la  10^{14.0}  \msunh$, indicate  that  the
luminosities  of central  galaxies are  clearly distinct  from those  of their
satellites.  The fraction of fossil groups, defined as those groups with $\log
L_1 -  \log L_2\ge  0.8$, ranges  from $\sim 2.5\%$  for groups  with $M_h\sim
10^{14}\msunh$ to 18-60\% for groups with $M_h\sim 10^{13}\msunh$.  The number
distribution of satellite galaxies in groups of a given mass follows a Poisson
distribution,  in agreement  with  the occupation  statistics  of dark  matter
sub-halos.  This provides strong support  for the standard lore that satellite
galaxies reside in sub-halos.  Finally, we measure the fraction of satellites,
which  changes  from $\sim  5.0\%$  for  galaxies  with $\rmag\sim  -22.0$  to
$\sim40\%$ for galaxies with $\rmag\sim -17.0$.
\end{abstract}




\section{Introduction}

In recent  years, the halo occupation distribution  and conditional luminosity
function have become  powerful statistical measures to probe  the link between
galaxies  and their  hosting dark  matter halos.   Although  these statistical
measures themselves do not give physical explanations of how galaxies form and
evolve, they provide important  constraints on various physical processes that
govern  the  formation  and  evolution  of  galaxies,  such  as  gravitational
instability,  gas  cooling,  star  formation,  merging,  tidal  stripping  and
heating, and  a variety of  feedback processes. In particular,  they constrain
how their efficiencies scale with halo mass.

The halo occupation distribution (hereafter  HOD), $P(N \vert M)$, which gives
the probability of finding $N$  galaxies (with some specified properties) in a
halo of mass  $M$, has been extensively used to  study the galaxy distribution
in dark matter halos and galaxy  clustering on large scales (e.g.  Jing, Mo \&
B\"orner 1998;  Peacock \&  Smith 2000; Seljak  2000; Scoccimarro  \etal 2001;
Jing, B\"orner  \& Suto 2002; Berlind  \& Weinberg 2002;  Bullock, Wechsler \&
Somerville  2002; Scranton  2002; Kang  \etal 2002;  Marinoni \&  Hudson 2002;
Zheng \etal 2002;  Magliocchetti \& Porciani 2003; Berlind  \etal 2003; Zehavi
\etal  2004, 2005;  Zheng \etal  2005;  Tinker \etal  2005).  The  conditional
luminosity function (CLF),  $\Phi(L \vert M) {\rm d}L$,  which refines the HOD
statistic by considering the average number of galaxies with luminosity $L \pm
{\rm  d}L/2$ that reside  in a  halo of  mass $M$,  has also  been extensively
investigated (Yang, Mo \& van den Bosch  2003; van den Bosch, Yang \& Mo 2003;
Vale \& Ostriker 2004, 2006; Cooray 2006; van den Bosch et al.  2007a) and has
been  applied to  various  redshift surveys,  such  as the  2dFGRS, the  Sloan
Digital Sky Survey  (SDSS) and DEEP2 (e.g.  Yan, Madgwick  \& White 2003; Yang
\etal 2004; Mo et al. 2004; Wang  \etal 2004; Zehavi \etal 2005; Yan, White \&
Coil  2004).   These  investigations  demonstrate  that  the  halo  occupation
statistics  are very  useful  in establishing  and  describing the  connection
between galaxies and  dark matter halos. Furthermore, they  also indicate that
the galaxy/dark halo connection can provide important constraints on cosmology
(e.g.,van den  Bosch, Mo \& Yang  2003; Zheng \& Weinberg  2007). Finally, the
HOD/CLF framework also  allows one to split the  galaxy population in centrals
and  satellites, and  to describe  their properties  separately  (e.g.  Cooray
2005; White \etal 2007; Zheng \etal 2007).

As has been pointed out in Yang et al.  (2005c; hereafter Y05c), a shortcoming
of  the  HOD/CLF  models  is   that  the  results  are  not  completely  model
independent. Typically,  assumptions have to be made  regarding the functional
form of either $P(N \vert M)$  or $\Phi(L \vert M)$.  Moreover, in all HOD/CLF
studies  to date,  the occupation  distributions  have been  determined in  an
indirect  way:  the  free  parameters  of  the  assumed  functional  form  are
constrained  using {\it  statistical}  data on  the  abundance and  clustering
properties  of  the  galaxy  population.   One may  hope  to  circumvent  this
shortcoming by directly measure  the dark matter distribution around galaxies.
Such measurements  can in principle be obtained  through gravitational lensing
and X-ray observations.  However, both methods are hampered by requirements on
the data quality and uncertainties  related to the interpretation of the data.
For instance, weak lensing  measurements, which requires high-quality imaging,
typically needs  to resort to the stacking  of many lens galaxies  in order to
get  a   detectable  signal,  but  this  stacking   severely  complicates  the
interpretation in terms of the halo  masses of the lens galaxies.  In the case
of X-ray  observations, robust  constraints can only  be obtained  for massive
clusters, but even here the interpretation  of the data can be complicated due
to the  presence of substructure and deviations  from hydrostatic equilibrium.
An alternative method  to directly probe the galaxy -  dark halo connection is
to use galaxy groups as a representation of dark matter halos and to study how
the galaxy population  changes with the properties of  the groups (e.g., Y05c;
Zandivarez et al. 2006; Robotham et al. 2006; Hansen \etal 2007).

Recently,  we have constructed  a large  galaxy group  catalogue based  on the
Sloan  Digital  Sky  Survey Data  Release  4  (SDSS  DR4), using  an  adaptive
halo-based group finder  (Yang \etal 2007; Paper I;  Y07 hereafter).  Detailed
tests with  mock galaxy catalogues have  shown that this group  finder is very
successful  in associating  galaxies  according to  their  common dark  matter
halos.  In  particular, the group finder  performs reliably not  only for rich
systems, but also for poor systems, including isolated central galaxies in low
mass halos.  This makes  it possible to  study the galaxy-halo  connection for
systems  covering a  large  dynamic range  in  masses.  Various  observational
selection effects  have been  taken into account,  especially the  survey edge
effects and  fiber collisions.  The halo  masses for the  groups are estimated
according to  the abundance match,  using the characteristic  group luminosity
and stellar masses (see \S\ref{sec_data} below).  According to tests with mock
galaxy catalogues, the halo masses  are estimated with a standard deviation of
about 0.3 dex.   With these well-defined galaxy group  catalogues, one can not
only study  the properties of galaxies  in different groups  (e.g.  Y05c; Yang
\etal 2005d; Collister \& Lahav 2005; van den Bosch \etal 2005; Robotham \etal
2006;  Zandivarez \etal  2006; Weinmann  \etal  2006a,b; van  den Bosch  \etal
2007b; McIntosh  \etal 2007), but also  probe how dark matter  halos trace the
large-scale  structure of  the Universe  (e.g.  Yang  \etal 2005b,  2006; Coil
\etal 2006;  Berlind \etal 2007; Wang  et al.  2007 in  preparation).  In this
paper, which is the second in a series, we use the SDSS DR4 group catalogue to
probe  various  occupation  statistics  and  measure the  CLFs  for  different
populations of galaxies.

This paper is organized as  follows: In Section~\ref{sec_data} we describe the
data (galaxy and group catalogues) used in this paper.  Section~\ref{sec_CLFs}
presents  our  measurement  of  the  CLFs  for all,  red  and  blue  galaxies.
Sections~\ref{sec_central}, ~\ref{sec_HOD} and ~\ref{sec_satfrac} describe the
properties  of  central  galaxies,  the  halo occupation  statistics  and  the
fraction  of  satellite galaxies,  respectively.   Finally,  we summarize  our
results  in  Section~\ref{sec_summary}.   Throughout  this  paper,  we  use  a
$\Lambda$CDM `concordance' cosmology whose  parameters are consistent with the
three-year   data  release   of  the   WMAP  mission:   $\Omega_m   =  0.238$,
$\Omega_{\Lambda}=0.762$,  $n_s=0.951$, $h=0.73$ and  $\sigma_8=0.75$ (Spergel
et al. 2007).

\begin{figure} \plotone{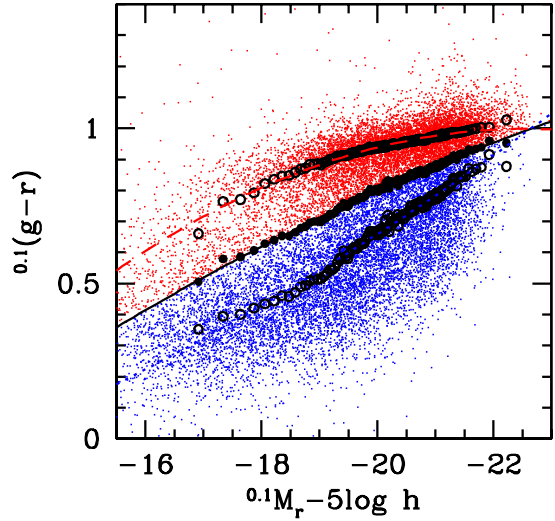}
  \caption{The color-magnitude relation for galaxies in our group sample.  The
    open circles indicate the Gaussian  peaks of the bi-normal distribution of
    galaxies in each luminosity bin. The solid dots indicate the corresponding
    averages  of the  two  Gaussian peaks.   The  solid line  is the  best-fit
    quadratic relation  to these averages (see  eq.~[\ref{quadfit}]), which we
    use  to split  the  galaxies  into red  and  blue population  (color-coded
    accordingly).}
\label{fig:data_color}
\end{figure}
\begin{figure*} \plotone{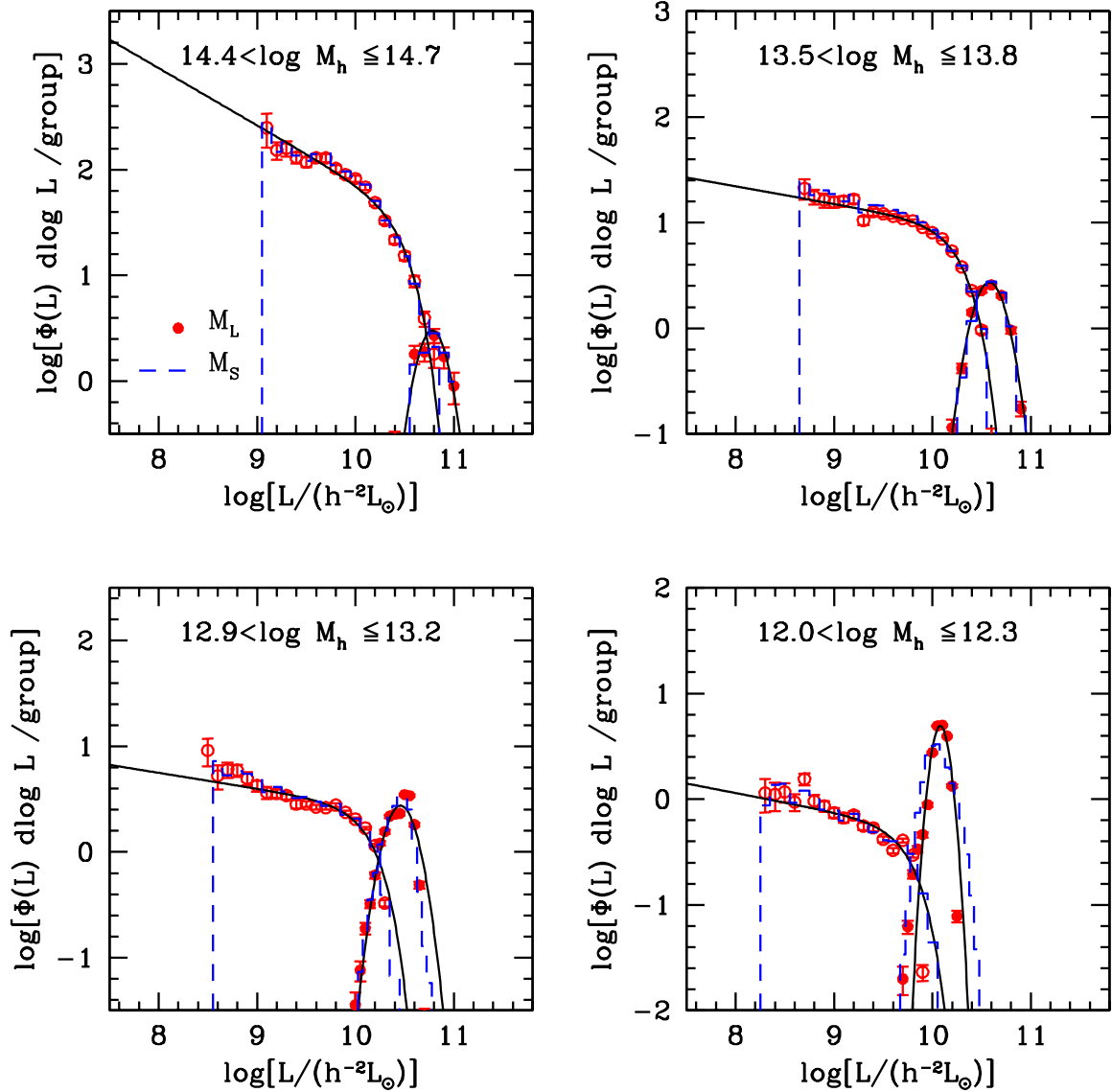}
  \caption{The conditional  luminosity functions (CLFs) of  galaxies in groups
    of different  mass bins.   Symbols correspond to  the CLFs  obtained using
    $M_L$  as   halo  mass  (estimated   according  to  the  ranking   of  the
    characteristic group  luminosity), with solid and  open circles indicating
    the contributions  from central and satellite  galaxies, respectively. The
    errorbars  reflect  the 1-$\sigma$  scatter  obtained  from 200  bootstrap
    samples.  The solid lines  indicate the related best-fit parameterizations
    using  equation~[\ref{eq:CLF_fit}].  For  comparison, we  also  show, with
    dashed  lines, the  CLFs  obtained  using $M_S$  as  halo mass  (estimated
    according  to the  ranking of  the group's  characteristic  stellar mass).
    Results shown in this plot are obtained from Sample II. }
\label{fig:CLF}
\end{figure*}
\begin{figure*} \plotone{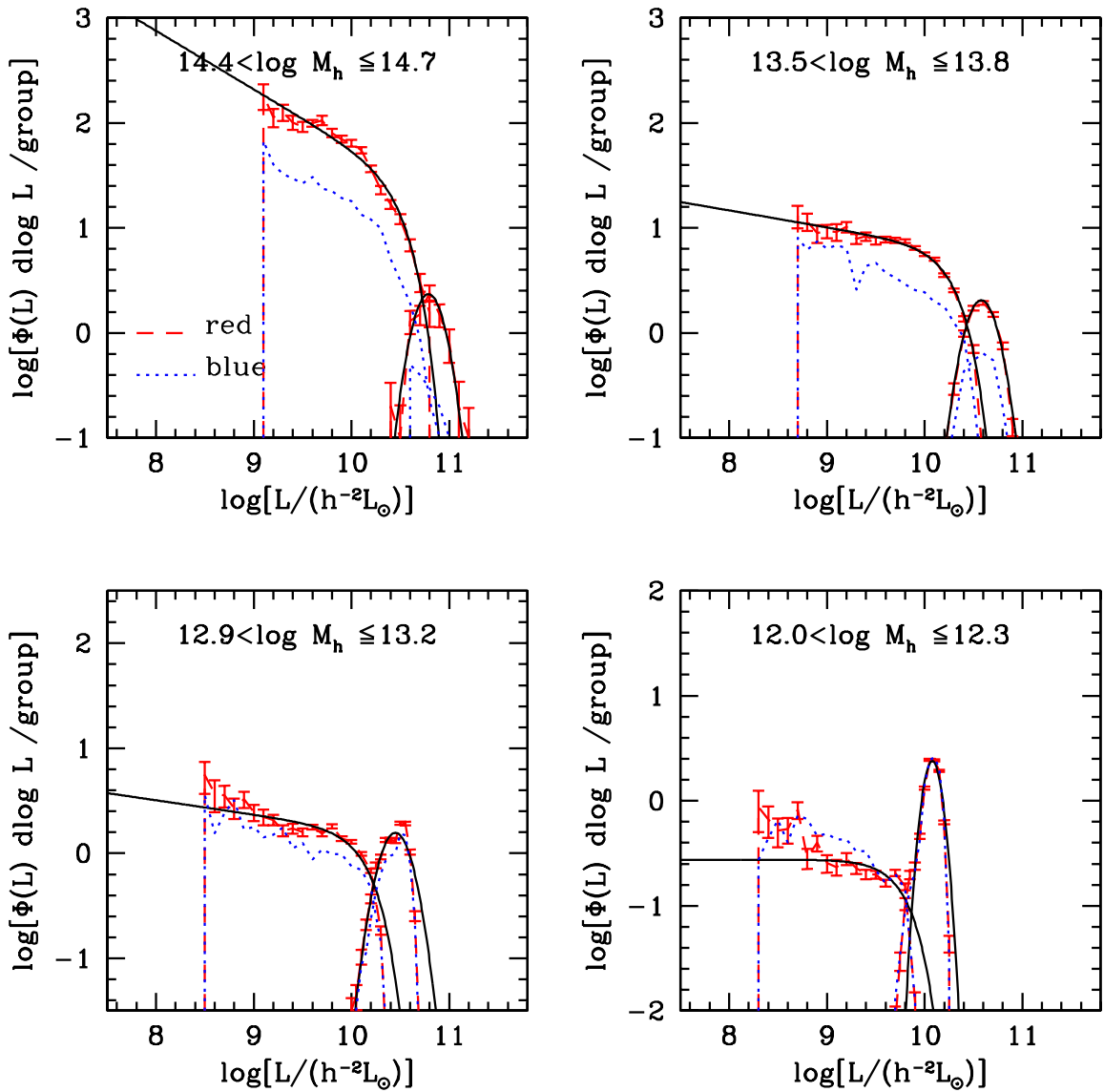}
  \caption{Similar to  Fig.~\ref{fig:CLF}, but here  we show the CLFs  for red
    (dashed  lines) and  blue (dotted  lines)  galaxies, for groups with  halo
    masses $M_L$. The solid  lines indicate the best-fit parameterizations for
    the  CLFs  of red  galaxies.   In both  cases  the  central and  satellite
    components  of  the  CLFs  are  indicated separately.   For  clarity,  the
    errorbars, again obtained using 200  bootstrap samples, are only shown for
    the red galaxies.  }
\label{fig:CLF_color}
\end{figure*}
\begin{figure*} \plotone{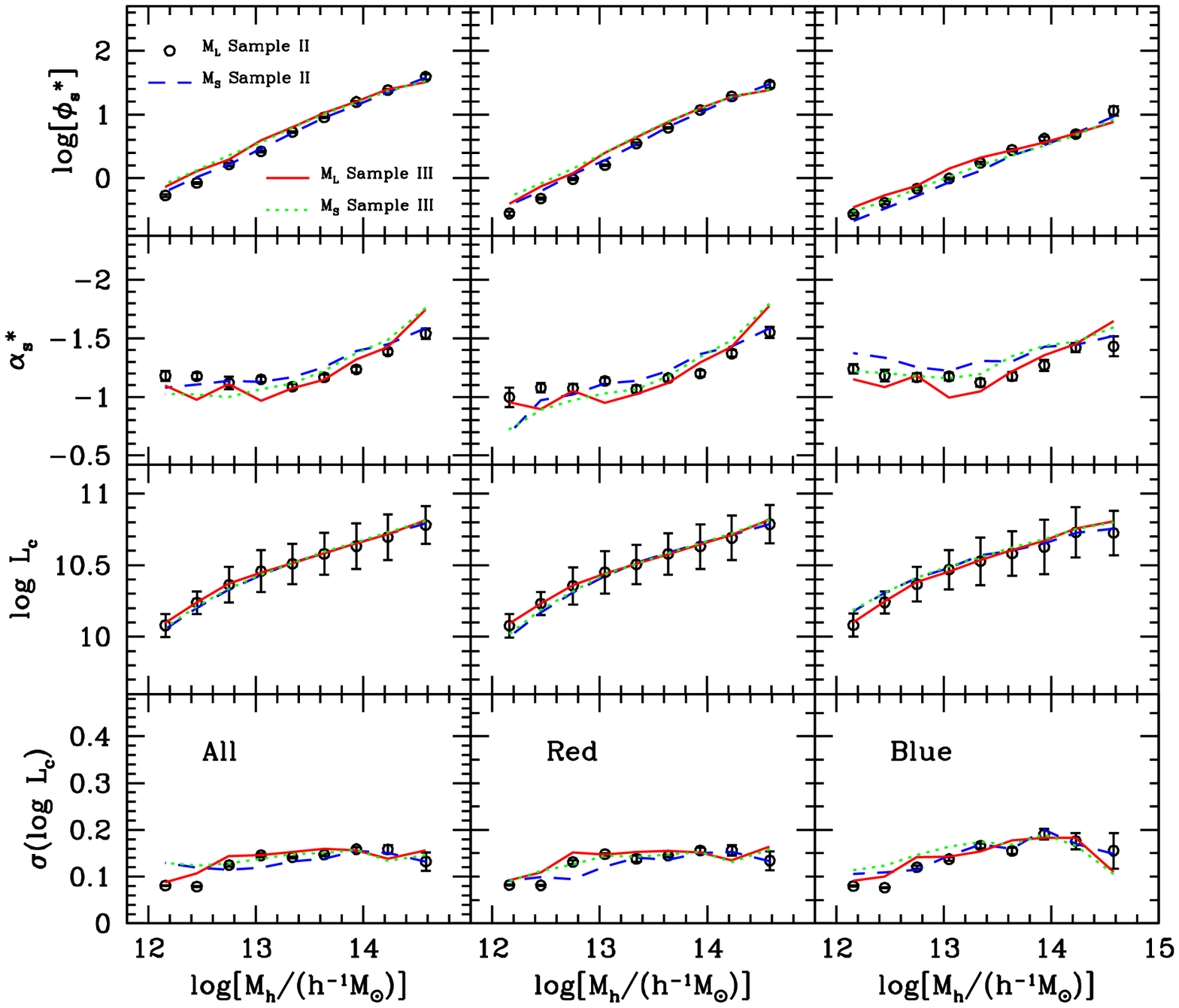}
  \caption{The    best   fit    parameters   ($\phi_s^{\star}$    upper   row,
    $\alpha_s^{\star}$ second row, $L_c$ third row, and $\sigma_c$ bottom row)
    to  the CLFs  shown  in Figs.~\ref{fig:CLF}  and ~\ref{fig:CLF_color},  as
    functions of  halo mass.  Panels  on the left,  in the middle, and  on the
    right show results for all, red, and blue galaxies, respectively. Since we
    have two  different halo  mass estimators ($M_L$  and $M_S$) and  two main
    group  samples (II  and III),  we have  obtained CLFs  for  four different
    combinations of sample and group mass estimator.  The results for all four
    combinations  are  shown  using  different  symbols  and  line-styles,  as
    indicated.  The errorbars  in the  first two  and last  rows  indicate the
    1-$\sigma$  variances obtained  from our  200 bootstrap  samples.   In the
    third row of  panels, however, the errorbars correspond  to the log-normal
    scatter, $\sigma_c$, shown  in the bottom row of  panels.  For clarity the
    errorbars are only shown for the `$M_L$-Sample II' case, but they are very
    similar for the other four cases.  }
\label{fig:fit_CLF}
\end{figure*}

\section{Data}
\label{sec_data}

The analysis  presented in this  paper is based  on the SDSS DR4  galaxy group
catalogue constructed by  Y07 using an adaptive halo-based  group finder (Yang
et  al.  2005a).   The related  galaxy catalogue  is the  New  York University
Value-Added Galaxy  Catalogue (NYU-VAGC; Blanton \etal 2005b),  which is based
on  the  SDSS  Data Release  4  (Adelman-McCarthy  \etal  2006), but  with  an
independent set of significantly  improved reductions.  From this catalogue we
select all  galaxies in  the Main  Galaxy Sample with  redshifts in  the range
$0.01 \leq  z \leq 0.20$ and with  a redshift completeness $\calC  > 0.7$.  As
described in Y07, three group  samples were constructed from the corresponding
galaxy samples: Sample I, which  only uses the $362356$ galaxies with measured
$r$-band magnitudes and redshifts from the SDSS, Sample II which also includes
7091  galaxies  with  SDSS   $r$-band  magnitudes  but  redshifts  taken  from
alternative  surveys,  and  Sample  III  which includes  an  additional  38672
galaxies that lack redshifts due to fiber collisions but that are assigned the
redshifts  of their  nearest neighbors  (cf.   Zehavi et  al.  2002).   Unless
stated otherwise, our analysis is based  on Sample II. For comparison, we also
present some results obtained from Sample III.

The  magnitudes and  colors of  all galaxies  are based  on the  standard SDSS
Petrosian technique (Petrosian 1976;  Strauss \etal 2002), have been corrected
for galactic  extinction (Schlegel, Finkbeiner  \& Davis 1998), and  have been
$K$-corrected and  evolution corrected to $z=0.1$, using  the method described
in Blanton \etal (2003a, b).  We  use the notation $^{0.1}M_r$ to indicate the
resulting  absolute magnitude in  the $r$-band.  Stellar masses,  indicated by
$M_*$,  for all  galaxies are  computed  using the  relations between  stellar
mass-to-light ratio and  $^{0.1}(g-r)$ color of Bell \etal  (2003; see Y07 for
details).

In this study  we separate galaxies into red and  blue subsamples according to
their bi-normal  distribution in the  $^{0.1}(g-r)$ color (Baldry  \etal 2004;
Blanton  \etal 2005a;  Li  \etal 2006).   Fig.~\ref{fig:data_color} shows  the
color-magnitude distribution of the galaxies  in our Sample II (dots) together
with  the two  peak  values of  the  bi-normal distribution  in each  absolute
magnitude bin (open circles)  (Cheng Li; private communication).  The galaxies
are separated into red and blue  subsamples using the solid line, which is the
best fit to the average of the two peak values in each absolute magnitude bin:
\begin{equation}\label{quadfit} 
^{0.1}(g-r) = 1.022-0.0651x-0.00311x^2\,,
\end{equation}
where $x=\rmag + 23.0$.

For each group in our catalogue we  have two estimates of its dark matter halo
mass $M_h$:  (1) $M_L$, which  is based on  the ranking of  the characteristic
group luminosity $L_{19.5}$ , and (2)  $M_S$, which is based on the ranking of
the     characteristic    group     stellar     mass    $M_{\rm     stellar}$,
respectively\footnote{$L_{19.5}$ and $M_{\rm  stellar}$ are, respectively, the
total luminosity and total stellar mass  of all group members with $\rmag \leq
-19.5$.}.  As shown  in Y07, these two halo masses  agree reasonably well with
each other, with a scatter that  decreases from $\sim 0.1$ dex at the low-mass
end to $\sim  0.05$ dex at the massive end.  Detailed  tests using mock galaxy
redshift surveys  have demonstrated that  the group masses thus  estimated can
recover the  true halo masses with  a 1-$\sigma$ deviation of  $\sim 0.3$ dex,
and are  more reliable than  those based on  the velocity dispersion  of group
members (Y05c; Weinmann \etal 2006;  Berlind \etal 2006; Y07).  Note also that
survey  edge effects  have been  taken into  account in  our  group catalogue:
groups that suffer severely from edge  effects (about 1.6\% of the total) have
been removed from the catalogue.  In  most cases, we take the brightest galaxy
in the group as the central galaxy (BCG) and all others as satellite galaxies.
In addition,  we also considered a case  in which the most  massive galaxy (in
terms of stellar  mass) in a group is considered as  the central galaxy (MCG).
Tests have  shown that for most  of what follows, these  two definitions yield
indistinguishable results.   Whenever the two definitions  lead to significant
differences, we present results for both.

\section{The Conditional Luminosity Functions for all, red and blue galaxies}
\label{sec_CLFs}

\begin{figure*} \plotone{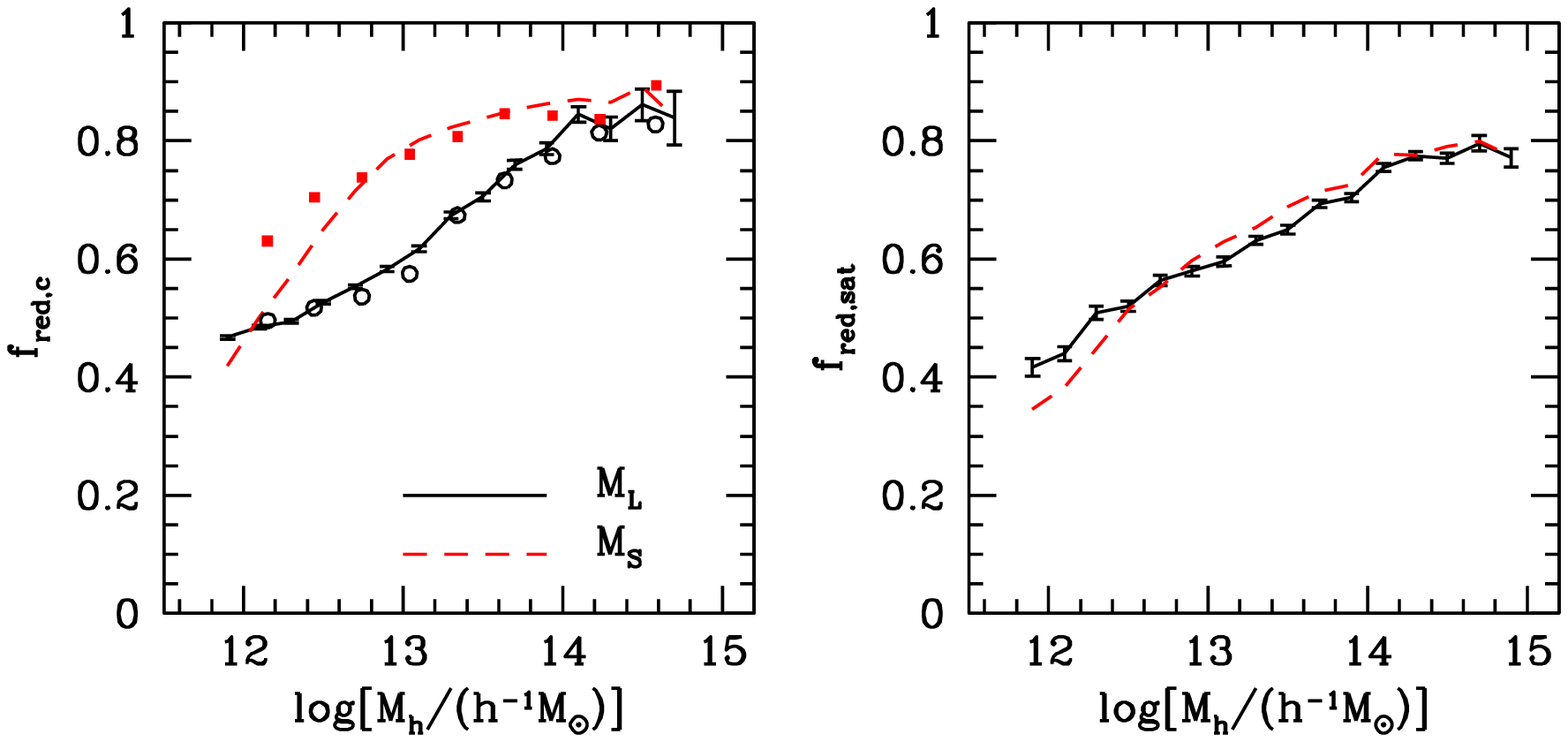}
  \caption{Fraction  of  red  galaxies  among central  (left-hand  panel)  and
satellite (right-hand  panel) galaxies.  Solid  and dashed lines  show results
for  groups with  halo mass  estimators  $M_L$ and  $M_S$, respectively.   For
comparison, we show  also the fractions of red  central galaxies obtained from
the CLFs  (Eq.~\ref{eq:phi_c}) as open  circles (for $M_L$) and  solid squares
(for  $M_S$).  Errorbars  (shown only  for $M_L$)  are obtained  from  the 200
bootstrap samples.  See text for a detailed discussion.}
\label{fig:f_red}
\end{figure*}

The conditional luminosity  function (CLF) of galaxies in  dark halos, $\Phi(L
\vert M)$,  which describes the  average number of  galaxies as a  function of
galaxy luminosity  in a dark matter halo  of a given mass,  plays an important
role in our understanding of how  galaxies form in dark matter halos (Yang, Mo
\& van den Bosch  2003; van den Bosch, Yang \& Mo;  2003; Cooray 2006; Vale \&
Ostriker 2006; van den Bosch \etal  2007a).  We now use our group catalogue to
directly determine $\Phi(L \vert M)$.

The CLF can be estimated by directly counting galaxies in groups.  For a given
galaxy  luminosity $L$,  there is  a  limiting redshift,  $z_L$, beyond  which
galaxies with such a luminosity are  not included in the sample.  As discussed
in Y07, the  group catalogue is complete to a certain  limiting redshift for a
given halo mass.  In  order to estimate the CLF, $\Phi(L \vert  M)$ at a given
$L$, we  only use groups that are  complete to the redshift  limit $z_L$.  The
CLF  is  obtained  by simply  counting  the  average  number of  galaxies  (in
luminosity bins) in groups of a  given $M$.  We show in Fig.~\ref{fig:CLF} the
resulting CLFs  for groups of  different masses. The contributions  of central
and  satellite  galaxies  are  plotted separately.   For  comparison,  results
obtained using  both $M_L$ and  $M_S$ are shown  as symbols and  dashed lines,
respectively.  Overall, these two  halo masses give consistent results, except
that the  $M_L$-based CLF of  the central galaxies  in low mass halos  is more
peaked  than  the $M_S$-based  CLF  (see  the  lower right-hand  panel).   The
errorbars shown in  each panel correspond to 1-$\sigma$  scatter obtained from
200  bootstrap  samples  of  our  group  catalogue.  The  CLF  for  the  total
population (centrals  plus satellites)  matches reasonably well  the Schechter
form down to halo masses  of $M\sim 10^{13.5}\msunh$.  For less massive halos,
however, there  is a prominent peak  in the CLF at  the bright end  due to the
contribution  of   central  galaxies,  which  makes  the   total  CLF  deviate
significantly  from the  Schechter form.   As discussed  in Y05c,  in low-mass
halos where  the group characteristic luminosity ($L_{19.5}$)  is dominated by
the central galaxy, the  $L_{19.5}$-$M_L$ conversion can produce an artificial
peak in  the CLF at the  bright end.  However, comparing  the results obtained
here with the  test results shown in Fig.~9 of Y05c  indicates that the strong
peak in the lower right panel  of Fig.~\ref{fig:CLF} cannot be entirely due to
the $L_{19.5}$-$M_L$  conversion.  Therefore, in  what follows, we  will model
the CLF for central and satellite galaxies separately.

Fig.~\ref{fig:CLF_color} shows  the CLFs separately for red (dashed lines) and
blue (dotted  lines) galaxies.  Note  that massive halos clearly  contain more
red  galaxies than  blue galaxies  (both centrals  and satellites),  while the
opposite applies  to low mass halos. The  overall CLF shapes for  red and blue
galaxies, however, are remarkably similar.

To quantify the CLFs, we fit each  of them with the following model.  We write
the total CLF as the sum of the CLFs of central and satellite galaxies:
\begin{equation}\label{eq:CLF_fit}
\Phi(L|M) = \Phi_{\rm cen}(L|M) + \Phi_{\rm sat}(L|M)\,.
\end{equation}
We assume the contribution from the central galaxies to be a lognormal:
\begin{equation}\label{eq:phi_c}
\Phi_{\rm cen}(L|M) = {A\over {\sqrt{2\pi}\sigma_c}} {\rm exp}
\left[- { {(\log L  -\log L_c )^2 } \over 2\sigma_c^2} \right]\,,
\end{equation}
where $A$ is  the number of central galaxies per halo.   Thus, $A\equiv 1$ for
all galaxies,  $A=f_{\rm red}$ (with $f_{\rm  red}$ being the  fraction of red
central  galaxies)   for  red  galaxies,   and  $A=1-f_{\rm  red}$   for  blue
galaxies. Note  that $\log L_c$ is,  by definition, the  expectation value for
the (10-based) logarithm of the luminosity of the central galaxy;
\begin{equation}
\log L_c = \int_0^{\infty} \Phi_{\rm cen}(L|M) \log L {\rm d}L\,,
\end{equation}
and that $\sigma_c=\sigma(\log L_c)$.  For the contribution from the satellite
galaxies we adopt a modified Schechter function:
\begin{equation}\label{eq:phi_s}  \Phi_{\rm  sat}(L|M)   =  \phi^*_s  \left  (
{L\over  L^*_s}\right  )^{(\alpha^*_s+1)}  {\rm  exp} \left[-  \left  ({L\over
L^*_s}\right )^2 \right].
\end{equation}
Note that  this function decreases faster  at the bright end  than a Schechter
function.   This CLF  parameterization has  a total  of five  free parameters:
$L_c$, $\sigma_c$, $\phi^*_s$, $\alpha^*_s$ and $L^*_s$.  In practice, we find
that  $\log L_c \sim  \log L^*_s  +0.25$ to  good approximation.  We therefore
adopt $\log L_c \equiv \log L^*_s +0.25$, throughout, which reduces the number
of free parameters to four.

For all  the CLFs  measured above,  the best fit  parameters, using  the model
described  by equation~\ref{eq:CLF_fit},  are shown  in Fig~\ref{fig:fit_CLF}.
Results are shown separately for  all (left column panels), red (middle column
panels) and blue (right column  panels) galaxies.  The error-bars in the upper
two  rows reflect  the  1-$\sigma$  scatter obtained  from  our 200  bootstrap
samples.   Here we also  compare results  obtained using  Samples II  and III.
Note  that  Sample II  does  not  include any  galaxies  missed  due to  fiber
collisions, while Sample  III includes all such galaxies  by assigning each of
them  the redshift  of its  nearest  neighbor. Although  this fiber  collision
correction works  well in roughly  60 percent of  all cases, the  remaining 40
percent  are assigned redshifts  that can  be very  different from  their true
values (Zehavi \etal 2002).  Samples II and III should therefore be considered
as two extremes as far as a treatment of fiber-collisions is concerned.  Given
that we also have  two kinds of halo masses, $M_L$ and  $M_S$, we have a total
of  four different combinations  for which  we have  determined the  CLF.  The
results  for all four  cases are  shown as  different symbols  in each  of the
panels of  Fig~\ref{fig:fit_CLF}.  As  an illustration of  how well  the model
fits the data, the  solid lines in Figs.~\ref{fig:CLF} and \ref{fig:CLF_color}
indicate the corresponding best-fit models.

The upper  row of Fig.~\ref{fig:fit_CLF}  shows the best fit  normalization of
the  CLF  for  satellite  galaxies,  which describes  the  average  number  of
satellite galaxies  with luminosity $\sim  L^*_s$ in a  group of a  given halo
mass.   As expected,  Sample III  gives  a higher  $\phi^*_s$, especially  for
low-mass groups.  Comparing  $\phi^*_s$ for red (upper middle  panel) and blue
(upper right  panel) galaxies,  one sees that  the fraction of  red satellites
increases with  halo mass.  The second row  shows the faint end  slopes of the
CLFs,  $\alpha^*_s$.  In  massive halos  with $M_{\rm  h}  \ga 10^{13}\msunh$,
$\alpha^*_s$  decreases (i.e.,  becomes  more negative)  with increasing  halo
mass,  both for  red and  for blue  galaxies.  In  halos with  $M_{\rm  h} \la
10^{13}\msunh$, however, $\alpha^*_s$ decreases  with increasing halo mass for
red satellites, but increases with  $M_{\rm h}$ for blue satellites, while the
faint-end slope for the entire satellite population (red and blue combined) is
roughly constant  at $\sim -1.1$.   The third row  of panels shows  that $\log
L_c$  increases with  halo  mass, for  both  red and  blue  centrals.  A  more
detailed discussion  regarding this and  other properties of  central galaxies
will  be presented in  Section~\ref{sec_central}).  Finally,  the last  row of
Fig~\ref{fig:fit_CLF}  shows  the  width  of  the log-normal  CLF  of  central
galaxies.  For the combined sample of red and blue galaxies we find an average
value of  $\sigma_c = \sigma(\log  L_c) \sim 0.15$  This is in  good agreement
with  constraints  obtained  by  Yang  \etal (2003)  and  Cooray  (2006)  from
clustering and  abundances of galaxies  in the 2dFGRS and  SDSS, respectively.
However, Zheng  et al.  (2007)  found, based on  HODs models for the  SDSS and
DEEP2, that the log-normal width  increases from $\sim 0.13$ for massive halos
with $M_h  \sim 10^{13.5}\msunh$ to  $\sim 0.3$ for  low mass halos  with $M_h
\sim 10^{11.5}\msunh$. This is quite different from our results, which suggest
that $\sigma(\log  L_c)$ {\it decreases}  with decreasing halo  mass. However,
this is  most likely an  artefact of  the method that  we used to  assign halo
masses to  our groups. As  mentioned above, our  halo masses are based  on the
ranking of either $L_{19.5}$ or  $M_{\rm stellar}$, which implies that we have
assumed a one-to-one  relation between halo mass and  these two indicators. At
the low  mass end, both  $L_{19.5}$ and $M_{\rm  stellar}$ are expected  to be
strongly  correlated with  the luminosity  of corresponding  central galaxies.
Therefore the values of $\sigma_c$ obtained here should be considered as lower
limits, in particular for low mass halos.

\begin{deluxetable*}{ccccccc}
  \tabletypesize{\scriptsize} 
  \tablecaption{The best fit parameters of CLFs for all, red and blue galaxies} 
  \tablewidth{0pt} 
  \tablehead{ Galaxy & $\log [M_{h}]$ & $\log \langle [M_{h}]\rangle$ &
  $\phi^*_s$ & $\alpha^*_s$ & $\log L_c$ & $\sigma_c$ \\ 
  \cline{1-7}\\ (1) & (2) & (3) & (4) & (5) & (6) & (7)} 
  
\startdata
& [14.40, 15.00) & $ 14.58$ & $ 35.51\pm  3.88$ & $ -1.66\pm  0.11$ & $10.799\pm 0.019$ & $ 0.141\pm 0.021$\\
& [14.10, 14.40) & $ 14.23$ & $ 23.68\pm  1.07$ & $ -1.44\pm  0.04$ & $10.714\pm 0.014$ & $ 0.146\pm 0.011$\\
& [13.80, 14.10) & $ 13.94$ & $ 15.27\pm  0.90$ & $ -1.33\pm  0.07$ & $10.649\pm 0.012$ & $ 0.157\pm 0.005$\\
& [13.50, 13.80) & $ 13.64$ & $  9.60\pm  0.82$ & $ -1.20\pm  0.05$ & $10.584\pm 0.006$ & $ 0.149\pm 0.009$\\
ALL & [13.20, 13.50) & $ 13.34$ & $  5.72\pm  0.60$ & $ -1.11\pm  0.04$ & $10.513\pm 0.003$ & $ 0.144\pm 0.009$\\
& [12.90, 13.20) & $ 13.05$ & $  3.27\pm  0.60$ & $ -1.08\pm  0.08$ & $10.442\pm 0.013$ & $ 0.137\pm 0.012$\\
& [12.60, 12.90) & $ 12.75$ & $  1.87\pm  0.29$ & $ -1.09\pm  0.06$ & $10.350\pm 0.020$ & $ 0.128\pm 0.012$\\
& [12.30, 12.60) & $ 12.45$ & $  1.12\pm  0.23$ & $ -1.07\pm  0.09$ & $10.224\pm 0.019$ & $ 0.107\pm 0.020$\\
& [12.00, 12.30) & $ 12.16$ & $  0.67\pm  0.12$ & $ -1.10\pm  0.06$ & $10.074\pm 0.018$ & $ 0.107\pm 0.026$\\
\cline{1-7}\\
& [14.40, 15.00) & $ 14.58$ & $ 27.14\pm  3.25$ & $ -1.68\pm  0.13$ & $10.801\pm 0.022$ & $ 0.147\pm 0.021$\\
& [14.10, 14.40) & $ 14.23$ & $ 18.72\pm  2.53$ & $ -1.43\pm  0.07$ & $10.709\pm 0.051$ & $ 0.144\pm 0.029$\\
& [13.80, 14.10) & $ 13.94$ & $ 11.79\pm  0.76$ & $ -1.30\pm  0.07$ & $10.644\pm 0.011$ & $ 0.153\pm 0.006$\\
& [13.50, 13.80) & $ 13.64$ & $  6.98\pm  0.80$ & $ -1.17\pm  0.04$ & $10.581\pm 0.005$ & $ 0.146\pm 0.008$\\
RED & [13.20, 13.50) & $ 13.34$ & $  3.97\pm  0.59$ & $ -1.07\pm  0.05$ & $10.510\pm 0.008$ & $ 0.144\pm 0.007$\\
& [12.90, 13.20) & $ 13.05$ & $  2.12\pm  0.44$ & $ -1.06\pm  0.09$ & $10.438\pm 0.011$ & $ 0.140\pm 0.012$\\
& [12.60, 12.90) & $ 12.75$ & $  1.18\pm  0.19$ & $ -1.03\pm  0.04$ & $10.335\pm 0.026$ & $ 0.126\pm 0.023$\\
& [12.30, 12.60) & $ 12.45$ & $  0.67\pm  0.15$ & $ -0.96\pm  0.09$ & $10.207\pm 0.031$ & $ 0.100\pm 0.014$\\
& [12.00, 12.30) & $ 12.16$ & $  0.39\pm  0.09$ & $ -0.84\pm  0.16$ & $10.046\pm 0.046$ & $ 0.089\pm 0.005$\\
\cline{1-7}\\
& [14.40, 15.00) & $ 14.58$ & $  9.21\pm  1.98$ & $ -1.55\pm  0.09$ & $10.770\pm 0.043$ & $ 0.130\pm 0.045$\\
& [14.10, 14.40) & $ 14.23$ & $  4.94\pm  0.57$ & $ -1.45\pm  0.05$ & $10.740\pm 0.024$ & $ 0.174\pm 0.019$\\
& [13.80, 14.10) & $ 13.94$ & $  3.59\pm  0.42$ & $ -1.37\pm  0.08$ & $10.658\pm 0.024$ & $ 0.191\pm 0.012$\\
& [13.50, 13.80) & $ 13.64$ & $  2.51\pm  0.27$ & $ -1.26\pm  0.08$ & $10.603\pm 0.017$ & $ 0.165\pm 0.011$\\
BLUE & [13.20, 13.50) & $ 13.34$ & $  1.69\pm  0.34$ & $ -1.17\pm  0.11$ & $10.546\pm 0.019$ & $ 0.166\pm 0.009$\\
& [12.90, 13.20) & $ 13.05$ & $  1.08\pm  0.24$ & $ -1.14\pm  0.10$ & $10.473\pm 0.016$ & $ 0.147\pm 0.011$\\
& [12.60, 12.90) & $ 12.75$ & $  0.66\pm  0.10$ & $ -1.20\pm  0.06$ & $10.392\pm 0.022$ & $ 0.131\pm 0.015$\\
& [12.30, 12.60) & $ 12.45$ & $  0.43\pm  0.08$ & $ -1.20\pm  0.10$ & $10.277\pm 0.037$ & $ 0.102\pm 0.020$\\
& [12.00, 12.30) & $ 12.16$ & $  0.28\pm  0.06$ & $ -1.25\pm  0.09$ & $10.135\pm 0.052$ & $ 0.098\pm 0.015$\\
\enddata

\tablecomments{Column (1): Galaxy sample.  Column (2): halo mass range. Column
  (3): average of the logarithm of  the halo mass.  Column (4)-(7): average of
  the best fit free parameters to  the four measurements of the CLFs, as shown
  in  Fig.  \ref{fig:fit_CLF}.  The  errors indicate  the scatter  among these
  four measurements  or the scatter  obtained from the 200  bootstrap samples,
  whichever is larger.}\label{tab:CLF}
\end{deluxetable*}

For reference, Table~\ref{tab:CLF}  lists the {\it average} values  of the CLF
fitting parameters obtained  from the four combinations of  Samples II and III
and group  masses $M_L$ and $M_S$.   The errorbars indicate  the scatter among
these  four  combinations or  the  scatter  obtained  from the  200  bootstrap
samples, whichever is larger.

As shown in  Weinmann \etal (2006b) and Baldry \etal (2006),  the red and blue
fractions of galaxies  as function of halo mass  provide important constraints
for models of galaxy  formation.  Fig.~\ref{fig:f_red} shows the red fractions
of centrals (left-hand panel) and satellites (right-hand panel) as function of
halo mass  obtained from our group  catalogue (solid and  dashed lines). These
fractions  are  obtained  by  simply  dividing  the  number  of  red  centrals
(satellites) by the  number of all centrals (satellites) in  a given halo mass
bin.   For  both centrals  and  satellites  the  red fraction  increases  with
increasing halo mass.  For comparison, results are shown for both $M_L$ (solid
lines) and $M_S$ (dashed lines).  These two different halo mass estimates give
quite different results  for central galaxies, especially for  halos with $M_h
\sim  10^{13}\msunh$.  The  origin of  this discrepancy  can be  understood in
terms of  Fig.  ~\ref{fig:data_color}.   In this color-magnitude  diagram, the
galaxies are  separated into  red and blue  populations.  If we  would convert
this color-magnitude  diagram into a  color-stellar mass diagram, the  red and
blue populations would shift slightly towards higher and lower stellar masses,
respectively  (according to the  Eq. 2  in Y07).   Therefore, a  certain upper
percentile of  galaxies that is ranked  according to stellar mass  will have a
larger red fraction  than the same percentile ranked  according to luminosity.
Because of the tight correlation  between the luminosity of the central galaxy
and $M_L$, and between the stellar mass of the central galaxy and $M_S$, it is
clear that centrals  in a given bin  of $M_S$ have a larger  red fraction than
those in the same bin of $M_L$.

We can  also determine  the red  fractions of centrals  from our  best-fit CLF
parameterizations   (equation~[\ref{eq:phi_c}]).     As   shown   above,   the
parameterization  of  $\Phi_{\rm  cen}(L|M)$  involves the  parameter  $f_{\rm
red}$, which  describes the  fraction of central  galaxies that are  red.  For
comparison, the  symbols in the  left-hand panel of  Fig.~\ref{fig:f_red} show
the best-fit  values of  $f_{\rm red}$  obtained from the  CLFs for  the $M_L$
masses (open circles) and $M_S$ masses (solid squares).  It is reassuring that
these best  fit results  match the direct  determination of the  red fractions
reasonably well.  The small differences between the direct measurement (lines)
and CLF measurement (symbols) owe to the  fact that in the latter case we have
normalized the galaxies  using the number of groups  within the redshift limit
$z_L$ (see Section~\ref{sec_CLFs}).

The  results shown  in the  left panel  indicate that  more than  80\%  of the
central  galaxies in  halos more  massive  than $\sim  10^{14}\msunh$ are  red
galaxies, while in  smaller halos with masses $\sim  10^{12}\msunh$, less than
50\% of  the centrals are  red. As discussed  in Y07, $M_S$ may  represent the
true  halo mass  better than  $M_L$.   The result  based on  $M_S$ shows  that
$f_{\rm red}$  decreases rapidly with decreasing  halo mass at  $M_{\rm h} \la
10^{13}\msunh$.  The right panel of Fig.~\ref{fig:f_red} shows the fraction of
red  satellite galaxies  as a  function of  halo mass.   As one  can  see, for
satellite galaxies $f_{\rm  red}$ increase steadily from about  40\% for halos
with $M_{\rm h}\sim 10^{12}\msunh$ to about 80\% in massive halos with $M_{\rm
h}  \sim 10^{15}\msunh$.  In van  den Bosch  \etal (2007b)  we have  used this
information  to constrain  the efficiency  with  which the  star formation  of
galaxies is quenched once they become a satellite galaxy (i.e., after they are
accreted into a larger halo).

\section{The properties of central galaxies}
\label{sec_central}
\begin{figure*} \plotone{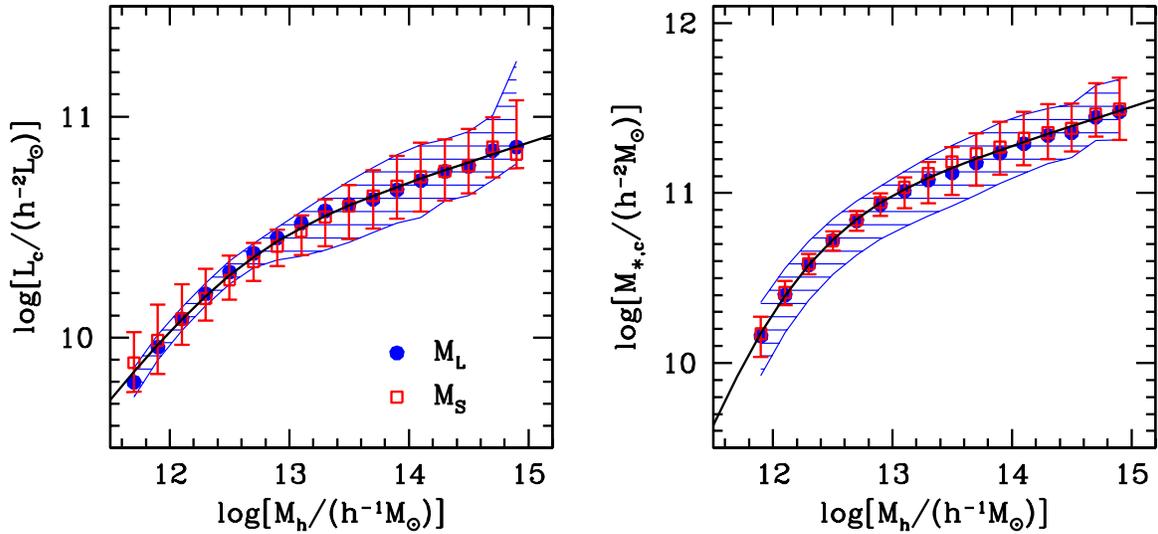}
  \caption{The left-hand panel shows the  median luminosity of the BCG, $L_c$,
    as function of halo mass, $M_h$.  The solid dots and open squares show the
    results for halo  mass $M_L$ and $M_S$, respectively.   The related shaded
    areas (or  error-bars) show the  $68\%$ confidence regions of  $L_c$.  The
    right-hand panel shows  the median stellar mass of  the MCG, $M_{*,c}$, as
    function of halo mass, $M_h$.  Again  solid dots and open squares show the
    results for  halo mass  $M_L$ and $M_S$,  and shaded area  (or error-bars)
    show the $68\%$ confidence regions of $M_{*,c}$.  The solid lines shown in
    the   left   and  right   panels   are   the   best  fit   results   using
    equations~\ref{eq:Lc_fit} and \ref{eq:Mc_fit}, respectively.}
\label{fig:L_c}
\end{figure*}

We now  turn to  a more  detailed investigation of  the properties  of central
galaxies in our group catalogue. We re-iterate that a central galaxy is either
defined as the  brightest group member (BCG) or the  most massive group member
(MCG).

\subsection{Central luminosity (stellar mass) - halo mass relation}
\label{sec:Lc}

In Fig.~\ref{fig:L_c},  we show  the luminosity -  halo mass (left  panel) and
stellar mass - halo mass (right panel) relations. The solid circles and shaded
areas  indicate the  median  and  68\% confidence  levels  of these  relations
obtained using $M_L$ as halo mass,  while the open squares with error-bars are
the results obtained using $M_S$ as halo mass.  Clearly, and not surprisingly,
the luminosity  (stellar mass) of the  BCG (MCG) increases with  halo mass. In
both cases we find that the slope of the relation decreases significantly with
increasing halo mass,  in good agreement with previous  results (e.g., Vale \&
Ostriker 2004,2006;  Cooray 2005; Yang \etal  2003; Y05c; van  den Bosch \etal
2007a). The physical reason for this change in slope is probably a combination
of  AGN feedback, and  changes in  the efficiencies  of radiative  cooling and
dynamical   friction  (e.g.    Lin   \etal  2004;   Dekel   2004;  Cooray   \&
Milosavljevi\'c 2005).

To quantify  the $L_c-M_h$ relation shown in  the left panel, we  fit the data
using the following function,
\begin{equation}\label{eq:Lc_fit}
L_c = L_0 \frac { (M_h/M_1)^{\alpha +\beta} }{(1+M_h/M_1)^\beta } \,.
\end{equation}
This model  contains four free  parameters: a normalized luminosity,  $L_0$, a
characteristic halo  mass, $M_1$,  and two slopes,  $\alpha$ and  $\beta$. The
solid line shown  in the left panel  is the best fit to  the average $L_c-M_h$
relation.  Note that  using $M_L$ or $M_S$  as the halo mass does  not lead to
any  significant changes  in the  results.   The best  fitting parameters  are
[$\log  L_0$, $\log  M_1$, $\alpha$,  $\beta$] =  [10.45, 12.54,  0.17, 0.51].
Thus,  according to Eq.~\ref{eq:Lc_fit},  $L_c$ scales  with $M_h$  roughly as
$L_c\propto  M_h^{0.17}$  for  halos  with $M_h\gg  10^{12.5}\msunh$,  and  as
$L_c\propto   M_h^{0.68}$    for   halos   with    $M_h\ll   10^{12.5}\msunh$.
Unfortunately,   since  we  do   not  have   data  for   halos  with   $M_h  <
10^{11.6}\msunh$,  any  change  of  behavior   at  the  low-mass  end  is  not
constrained. For the $M_{*,c}-M_h$ relation shown in the right panel, we use a
similar function to fit the data:
\begin{equation}\label{eq:Mc_fit}
M_{*,c} = M_0 \frac { (M_h/M_1)^{\alpha +\beta} }{(1+M_h/M_1)^\beta } \,.
\end{equation}
The solid line shown  in the right panel is the best fit  of this model to the
data,  and the  best-fit parameters  are  [$\log M_0$,  $\log M_1$,  $\alpha$,
$\beta$] = [10.86,  12.08, 0.22, 1.61].  Thus, $M_{*,c}$  scales with $M_h$ as
$M_{*,c}\propto M_h^{0.22}$  for halos  with $M_h\gg 10^{12.1}\msunh$,  and as
$M_{*,c}\propto M_h^{1.83}$ for halos with $M_h\ll 10^{12.1}\msunh$.

\begin{figure*} \plotone{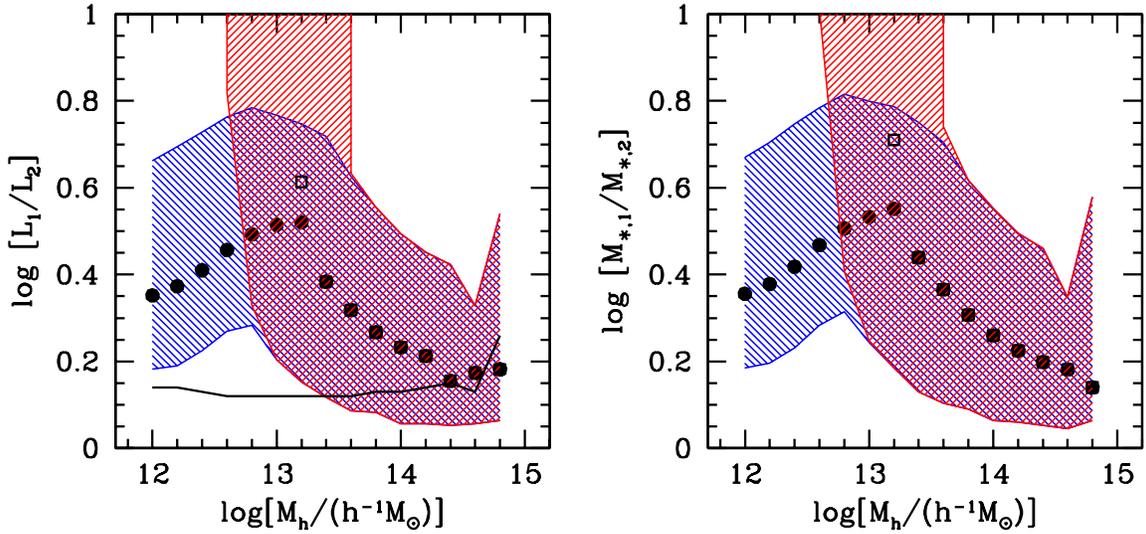}
  \caption{The left-hand  panel shows the  luminosity gap between the  BCG and
    the  brightest satellite  galaxy in  the  group, $\log  L_1-\log L_2$,  as
    function of group mass $M_h$.  The right-hand panel shows the stellar mass
    gap between  the MCG  and the most  massive satellite galaxy,  $\log M_{*,
      1}-\log M_{*,  2}$ as function of  group mass $M_h$.   Results are shown
    for two  cases, where groups with  only 1 member  are treated differently.
    In  case  1,  we assume  that  the  brightest  satellite galaxy  has  zero
    luminosity (open  squares), while in case  2 we assume  that the brightest
    satellite galaxy has an apparent magnitude equal to the magnitude limit of
    the  survey (solid  dots).  The  shaded areas  indicate  the corresponding
    $68\%$ confidence intervals.   The solid line in the  left panel shows the
    best fit values of $\sigma (\log L_c)$ obtained from the CLFs. }
\label{fig:L1L2}
\end{figure*}
\begin{figure*} \plotone{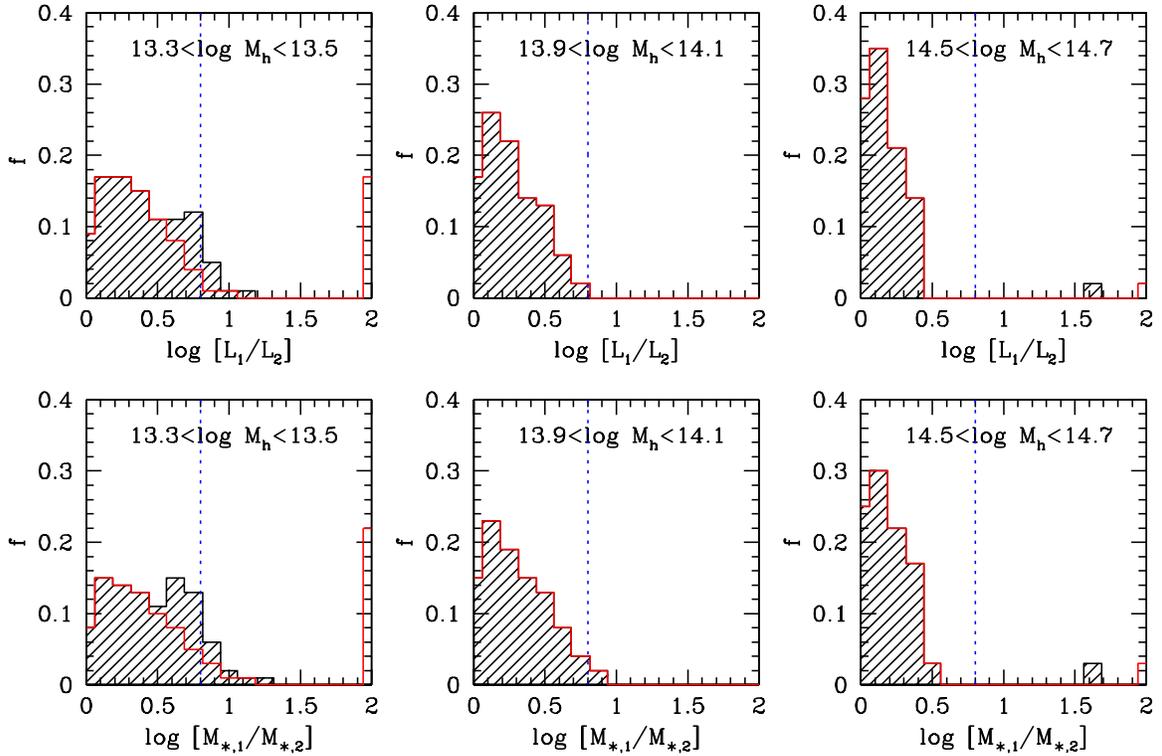}
  \caption{The  probability distributions  of luminosity  gap $\log  L_1 -\log
    L_2$ (upper  panels) and  stellar mass gap  $\log M_{*, 1}-\log  M_{*, 2}$
    (lower panels) in  halos of different mass ranges,  as indicated.  Similar
    to  Fig.~\ref{fig:L1L2},  results are  shown  for  both  case 1  (unshaded
    histograms)  and case  2  (shaded  histograms). In  case  1, the  isolated
    galaxies are put into the bin  with $\log L_1-\log L_2=2.0$ or $\log M_{*,
      1}-\log M_{*, 2}=2.0$.}
\label{fig:pL1L2}
\end{figure*}

\subsection{The luminosity (stellar mass)-gap statistic}
\label{sec:gap}

An  useful  quantity  to  describe  the difference  between  the  central  and
satellite galaxies is  the so-called `luminosity-gap' between the  BCG and the
brightest  satellite galaxy  in  a given  halo.   This `luminosity-gap'  holds
important information  regarding the formation and evolution  of galaxies. For
example,  as discussed  in  D'Onghia \etal  (2005)  and Milosavljevi\'c  \etal
(2006), the  luminosity-gap may  be used  to quantify the  dynamical age  of a
system of  galaxies: halos  with $L_2/L_1$ close  to unity must  be relatively
young, as  dynamical friction  causes multiple luminous  galaxies in  the same
halo  to  merge  on  a  relatively  short time  scale.  Put  differently,  the
distribution of $L_2/L_1$ holds important information regarding the importance
of galactic  cannibalism for BCGs (e.g.,  Tremaine \& Richstone  1977; Vale \&
Ostriker 2007)

Using our group catalogue, it is straightforward to measure $L_2/L_1$, as well
as the stellar mass equivalent,  $M_{*,2}/M_{*,1}$, as function of group mass,
as  long as  the group  has at  least two  members. In  the case  of `isolated
galaxies', i.e., groups with only one member (the central), however, there are
two  possibilities: either  the galaxy  is truly  isolated, in  that  its halo
contains no  satellite galaxies,  or the halo  contains one or  more satellite
galaxies that are  fainter than the flux limit of the  survey.  To bracket the
extremes we proceed  as follows.  For isolated galaxies  we either assume that
the brightest satellite has zero luminosity (case 1) or that its luminosity is
such that its apparent magnitude limit  is equal to the magnitude limit of our
sample (case 2).  Fig.~\ref{fig:L1L2} shows the results thus obtained for both
the  luminosity gap  (left-hand panel)  and the  stellar mass  gap (right-hand
panel).  Symbols  and shaded  areas indicate the  medians and  68\% confidence
intervals, respectively, and results are  shown for both case 1 (open squares)
and case 2 (solid dots).  For groups with $M_h\ga 10^{13.5}\msunh$, case 1 and
2 yield  identical results, simply because  all these groups  contain at least
two member  galaxies above  the magnitude  limit of the  survey.  In  the most
massive halos,  with $M_h\ga 10^{14.5}\msunh$, the  median gap is  $\log L_1 -
\log L_2 \sim \log M_1 - \log M_2 \sim 0.2$.  The median luminosity or stellar
mass  gap  can  be  reliably  measured  down to  a  halo  masses  of  $M_h\sim
10^{13.5}\msunh$, where the  median values are $\log L_1 -  \log L_2 \sim 0.3$
and $\log M_1 - \log M_2  \sim 0.4$.  For halos with $M_h\la 10^{13.5}\msunh$,
however, cases 1 and 2 yield  very different results, indicating that the flux
limit of the survey severely impedes our ability to accurately measure the gap
statistics.  Finally we emphasize  that the luminosity-gap statistics obtained
from Samples  II and  III separately are  very similar, indicating  that fiber
collisions do not have a strong impact on the results presented here.

As pointed  out by Tremaine \&  Richstone (1977; and  references therein), the
`luminosity-gap' can  be used to determine  whether the BCGs  in galaxy groups
are statistically drawn  from the same distribution as  the satellite galaxies
or whether  they are `special'.   If the average magnitude  difference between
the BCGs  and the  brightest satellite galaxies  is smaller than  the standard
deviation in the  magnitudes of the BCGs, then they  are consistent with being
draw  from the  same  distribution.  To  test  this, the  solid  curve in  the
left-hand  panel  of  Fig.~\ref{fig:L1L2}  indicates the  best-fit  values  of
$\sigma_c = \sigma(\log  L_c)$ obtained from the CLFs.   A comparison with the
median  luminosity gaps  suggests that  the  BCGs in  groups, especially  with
$M_h\la   10^{14.0}\msunh$,  form   a  `special'   subclass,  in   that  their
luminosities can not  be considered the extreme values  of the distribution of
satellite luminosities.  As a cautionary  remark, we emphasize that because of
the method used  to assign halo masses to the groups,  the value of $\sigma_c$
may  be underestimated,  especially for  low  mass groups  (see discussion  in
section~\ref{sec_CLFs}).  However, even if $\sigma_c$ were underestimated by a
factor two at the low mass end, this would not change our conclusion that BCGs
are special in low mass halos.

The `luminosity-gap' can  also be described using its  distribution for groups
of  a given  mass. In  the upper  panels of  Fig.~\ref{fig:pL1L2} we  show the
distribution of $\log L_1 -\log L_2$  for groups in the above mentioned case 1
(unshaded histograms)  and case 2 (shaded  histograms), respectively.  Results
are shown for  three different mass bins, as indicated in  the panels.  In the
lower panels of Fig.~\ref{fig:pL1L2},  results are shown for the corresponding
stellar mass  gap distributions.  One can  see, the width  of the distribution
increases with decreasing halo mass.

Systems with a relatively large luminosity  gap, which most likely owes to the
fact that  the brightest galaxies  in the halo  have merged, are  often termed
``fossil groups'' and  have received a significant amount  of attention in the
recent literature (see Vikhlinin \etal  1999; Jones \etal 2003; D'Onghia \etal
2005;  Milosavljevi\'c \etal  2006; Sommer-Larsen  2006; van  den  Bosch \etal
2007a; Sales  et al.  2007; von Benda-Beckmann  \etal 2007).   Following Jones
\etal (2003) and  Milosavljevi\'c \etal (2006) we define  systems in which the
brightest  satellite galaxy  is at  least 2  magnitudes fainter  than  the BCG
(i.e., $\log L_1 -\log L_2 \ge 0.8$, indicated as the dotted vertical lines in
Fig.~\ref{fig:pL1L2}),  as  ``fossil''  systems.   From  the  SDSS  DR4  group
catalogue, we obtain that the  fraction of fossil systems increases from $\sim
0.5$ percent for groups with $M_h \sim 10^{14.5}\msunh$, to $\sim 2.5$ percent
for groups  with $M_h \sim  10^{14.0}\msunh$, $11-20$ percent for  groups with
$M_h  \sim 10^{13.5}\msunh$,  and  $18-60$ percent  for  groups with  $M_h\sim
10^{13.0}\msunh$\footnote{Whenever  two values are  quoted, these  reflect the
two extreme cases described above.}.  These results are in good agreement with
a  similar analysis of  galaxy groups  in the  2dFGRS by  van den  Bosch \etal
(2007a). On the  other hand, Jones \etal (2003) obtained  an incidence rate of
$8$ to $20$ percent for systems with an X-ray luminosity from diffuse, hot gas
of  $L_{\rm X, bol}  \geq 2.5  \times 10^{41}  h^{-2} {\rm  erg}{\rm s}^{-1}$.
However, since the groups in our  SDSS DR4 catalogue are not X-ray selected, a
detailed  comparison with our  results is  not possible.   In a  recent paper,
D'Onghia \etal (2005) used  detailed hydrodynamical simulations to predict the
fraction of halos with $M_h \sim 10^{14} \msunh$ that have $\log L_1 -\log L_2
\ge  0.8$. From  a total  of  twelve simulated  groups, they  obtain a  fossil
fraction of $33 \pm 16$ percent.   This value is much higher than the fraction
of fossil systems we find in the SDSS, which suggests a potential over-merging
problem in  their simulations.  More recently von  Benda-Beckmann \etal (2007)
used  a  combination  of  N-body  simulations and  empirical  models  for  the
connection  between galaxy  luminosity and  halo  mass (taken  from Cooray  \&
Milosavljevi\'c )2005), and found that the fossil group fraction is about 24\%
among all  systems with masses  $1-5\times 10^{13.0}\msunh$.  This is  in good
agreement with our direct measurement from the SDSS data.
\begin{figure*} \plotone{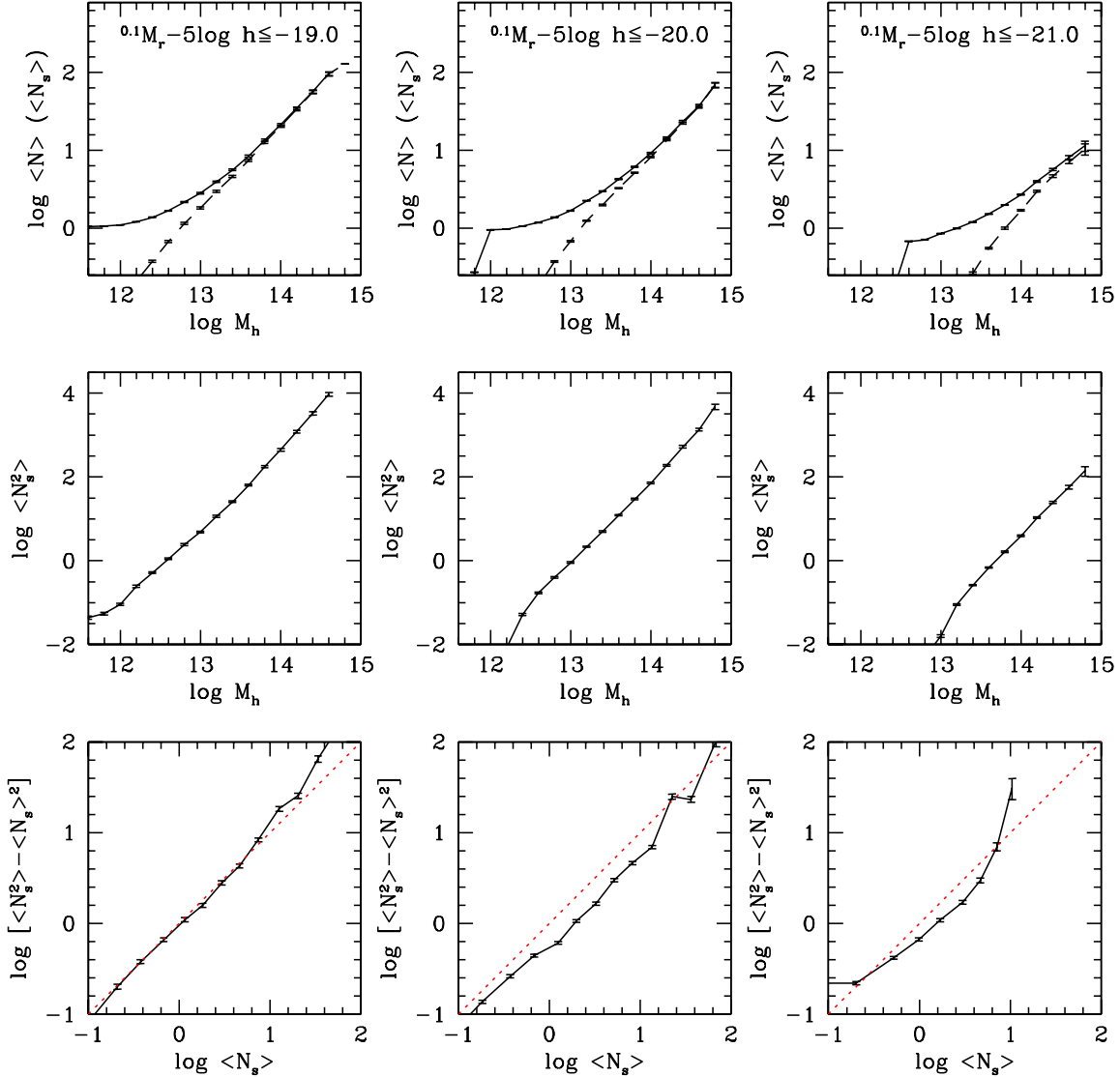}
  \caption{{\it Upper row  of panels:} the average halo  occupation numbers as
    function of halo mass for  all galaxies ($\langle N \rangle$, solid lines)
    and   for  satellite   galaxies  only   ($\langle  N_s   \rangle$,  dashed
    lines). Here  halo masses are taken to  be $M_L$.  Panels on  the left, in
    the middle and  on the right correspond to an  absolute magnitude limit of
    $\rmag  \le -19.0$,  $-20.0$  and $-21.0$,  respectively.  The  error-bars
    indicate   the   1-$\sigma$   variances   obtained  from   200   bootstrap
    samples. {\it Middle  row of panels:} similar to the  upper row of panels,
    except that  here we plot the  second moment of the  occupation numbers of
    satellite galaxies,  $\langle N_s^2 \rangle$.  {\it Lower  row of panels:}
    similar to  the middle  row of  panels, except that  now we  plot $\langle
    N_s^2 \rangle-\langle N_s \rangle^2$ as function of $\langle N_s \rangle$.
    The  dotted, diagonal  line indicates  $\langle N_s^2  \rangle-\langle N_s
    \rangle^2   =   \langle   N_s   \rangle$  and   corresponds   to   Poisson
    statistics. Apparently,  the distribution  of $N_s$ is  very similar  to a
    Poisson distribution.}
\label{fig:N_Mh}
\end{figure*}
\begin{figure*} \plotone{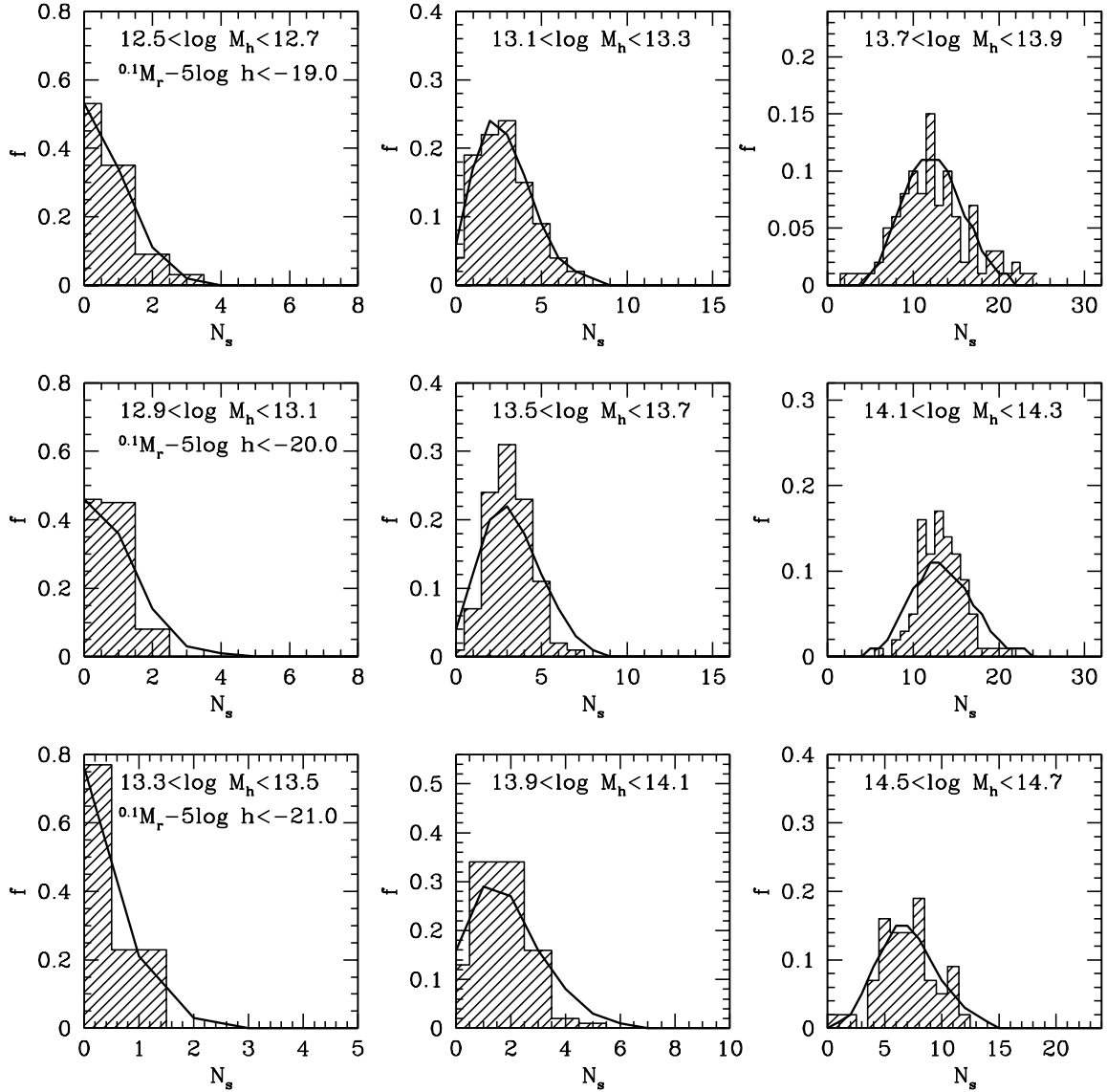}
  \caption{Number  distributions  of  the  satellite  galaxies  in  groups  of
    different halo mass  bins, as indicated.  Panels in  the upper, middle and
    lower rows correspond to different absolute magnitude limits as indicated.
    The hatched histograms indicate the distributions obtained from the groups
    in the SDSS.  Thick solid  curves correspond to Poisson distributions with
    the same mean $N_s$, and are  shown to illustrate the Poissonian nature of
    $P(N_s \vert M_h)$. }
\label{fig:Ns_Mh}
\end{figure*}

\section{Halo-occupation statistics}
\label{sec_HOD}

\begin{deluxetable}{ccc}
  \tabletypesize{\scriptsize} \tablecaption{Halo occupation of satellite galaxies}
  \tablewidth{0pt} \tablehead{ Satellite galaxies & $\log M_{s,0}$ & $\gamma$  \\
    \cline{1-3}\\
    (1) & (2) & (3)  } 
  
  \startdata
  $\le -18.0$ & $12.48\pm 0.04$ &  $1.01\pm 0.06$ \\
  $\le -18.5$ & $12.62\pm 0.02$ &  $1.05\pm 0.04$ \\
  $\le -19.0$ & $12.77\pm 0.02$ &  $1.06\pm 0.03$ \\
  $\le -19.5$ & $12.93\pm 0.01$ &  $1.07\pm 0.02$ \\
  $\le -20.0$ & $13.15\pm 0.01$ &  $1.09\pm 0.02$ \\
  $\le -20.5$ & $13.44\pm 0.01$ &  $1.10\pm 0.02$ \\
  $\le -21.0$ & $13.82\pm 0.01$ &  $1.13\pm 0.02$ \\
  $\le -21.5$ & $14.34\pm 0.01$ &  $1.33\pm 0.07$ \\
  \enddata
  
  \tablecomments{Halo  occupation model parameters  for satellite  galaxies in
    the SDSS  DR4. Here  the mean halo  occupations of satellite  galaxies are
    described by $\langle N_s \rangle =(M_h/M_{s,0})^\gamma$.  Column (1): The
    absolute  magnitude limit of  the satellite  galaxies. Columns  (2-3): the
    best fit  parameters $M_{s,0}$ and  $\gamma$ (averages with  errors).  The
    errors are estimated  from the variances between Samples  II and III, with
    halo masses $M_L$ and $M_S$, and  are much larger than the errors obtained
    from our 200 bootstrap samples.}\label{tab:HOS}
\end{deluxetable}

The upper panels of Fig.~\ref{fig:N_Mh} show the mean halo occupation numbers,
$\langle N  \rangle$ (the mean number  of all group members)  and $\langle N_s
\rangle$ (the  mean number  of satellites), as  functions of halo  mass $M_h$.
Results  are shown  only  for  $M_h=M_L$, but  adopting  $M_h=M_S$ gives  very
similar  results.  The  results shown  in the  left, middle  and  right panels
correspond to  galaxies in  different luminosity ranges,  as indicated  in the
panels.  Similar to what is found in Y05c, the sharp break at the low mass end
at $\langle  N \rangle \sim  1$ \footnote{This is  not seen in the  upper left
panel because  our group  catalogue does not  include halos with  masses below
$10^{11.6}\msunh$.}   indicates an almost  deterministic relation  between the
luminosity of  a central  galaxy and the  mass of  its dark matter  halo.  The
shoulder suggests  that the  brightest satellite galaxies  is in  general much
fainter than the central galaxy (e.g.  Zheng et al.  2005). The mean number of
satellite galaxies,  which is shown  as the dashed  line, reveals a  very good
power  law  feature.  To quantify  this.  we  model  the halo  occupation  for
satellite  galaxies with  $\langle N_s  \rangle  =(M_h/M_{s,0})^\gamma$, where
$M_{s,0}$ is  a characteristic halo  mass above which  there is on  average at
least one  satellite galaxies and $\gamma$  is the power law  index.  The best
fit parameters  for satellite galaxies with different  absolute magnitude cuts
are listed in  Table ~\ref{tab:HOS}. As an illustration, we  show in the upper
panels of Fig.~\ref{fig:N_Mh} the  corresponding best fit model predictions of
the halo occupation of satellite galaxies  as the dotted lines. They all agree
with the data extremely well. 

In addition to the mean halo occupation number, we also investigate the second
moment of the halo occupation  distribution (see Y05c).  Here we only consider
satellite galaxies, because by definition  a central galaxy is always assigned
to a group in a deterministic  way.  This quantity is crucial in modelling the
two-point correlation function of galaxies on small scales (e.g., Benson \etal
2000;  Berlind \etal  2003;  Y05c;  Tinker \etal  2007),  and holds  important
information regarding the physical  processes related to galaxy formation.  In
earlier investigations, a number of simple models were adopted to describe the
second moment of  the halo occupation distribution and  its dependence on halo
mass  (e.g., Berlind \&  Weinberg 2002;  Yang et  al.  2003).   In particular,
using the group samples constructed  from the 2dFGRS, Y05c measured the second
moment  for  {\it  all} group  members  and  found  that the  halo  occupation
distribution  is  close  to  Poissonian  in massive  halos  and  significantly
sub-Poissonian in low  mass halos.  In the middle  row of Fig.~\ref{fig:N_Mh},
we show $\langle N^2_s \rangle$ for  {\it satellite} galaxies as a function of
halo  mass  $M_h$.   As  one  can  see, $\langle  N^2_s  \rangle$  is  roughly
proportional to $M_h^2$.  To see  how the distribution deviates from a Poisson
distribution,  we show, in  the lower  panels of  Fig~\ref{fig:N_Mh}, $\langle
N^2_s \rangle - \langle N_s \rangle^2$ as a function of $\langle N_s \rangle$.
A Poisson  distribution has $\langle N^2_s  \rangle - \langle  N_s \rangle^2 =
\langle  N_s \rangle$ (doted  lines), while  a deterministic  distribution has
$\langle  N^2_s \rangle -  \langle N_s  \rangle^2 \sim  0$.  The  results thus
indicate that the number distribution  of satellite galaxies follows roughly a
Poisson distribution.

Fig.~\ref{fig:Ns_Mh}  shows  the  distribution  of  the  number  of  satellite
galaxies in  groups, $N_{s}$, for different  halo mass bins.   The thick solid
curves indicate  Poisson distributions with the same  $\langle N_{s} \rangle$.
As one can see, the  observed $N_{s}$-distributions are well fitted by Poisson
distributions.  These  properties, which  have  already  been  found in  Y05c,
suggest a  direct link between  satellite galaxies and dark  matter sub-halos.
In a recent  study, Kravtsov \etal (2004), using  large numerical simulations,
have shown that  the occupation distribution of dark  matter sub-halos follows
Poisson  statistics.  This suggests  that there  may be  a tight  link between
satellite  galaxies and  dark  matter  sub-halos, which  has  been a  standard
assumption in various HOD/CLF models  (e.g., Vale \& Ostriker 2004; 2006). Our
results provide observational support for such a link.

\section{Satellite Fractions}
\label{sec_satfrac}

\begin{figure*}
  \plotone{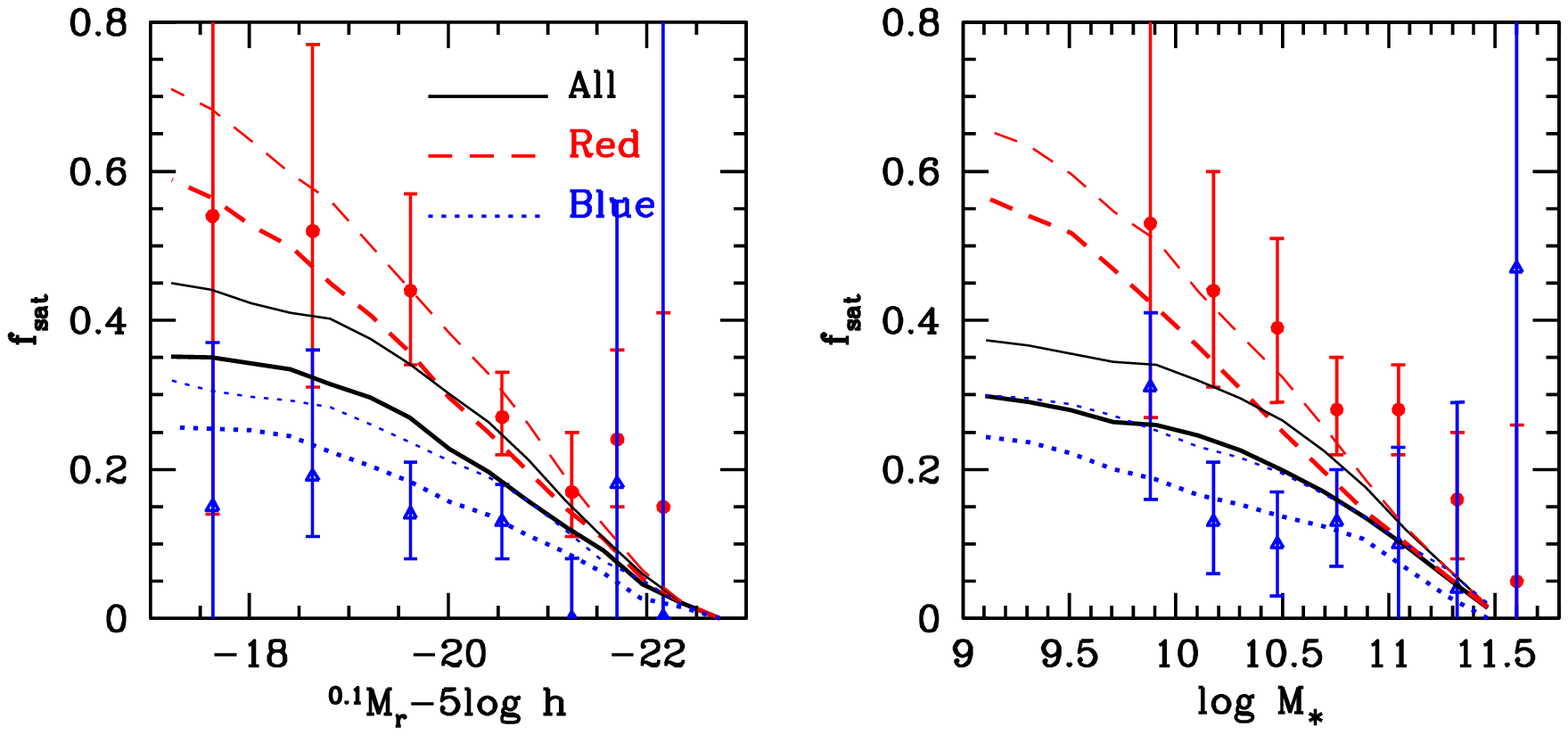}
  \caption{The  fraction  of  satellite  galaxies as  function  of  luminosity
    (left-hand panel) and stellar  mass (right-hand panel).  Results are shown
    separately for all,  red and blue galaxies as  indicated.  The thick lines
    are results for group Sample II  and thin lines for group Sample III.  For
    comparison, we  show the satellite fractions obtained  by Mandelbaum \etal
    (2006)  from a galaxy-galaxy  lensing analysis  of the  SDSS as  the solid
    circles and  open triangles with  vertical errorbars (95\% CL)  for early-
    and late-type galaxies, respectively}
\label{fig:f_sat}
\end{figure*}

\begin{deluxetable}{cccc}
  \tabletypesize{\scriptsize} \tablecaption{Satellite Fractions}
  \tablewidth{0pt} \tablehead{ Galaxy & All & Red & Blue  \\
    \cline{1-4}\\
    (1) & (2) & (3) & (4) } \startdata
  -22.34   &   0.021 $\pm$ 0.001   &   0.023 $\pm$ 0.001   &   0.017 $\pm$ 0.003\\
  -21.95   &   0.053 $\pm$ 0.010   &   0.058 $\pm$ 0.008   &   0.040 $\pm$ 0.018\\
  -21.56   &   0.100 $\pm$ 0.012   &   0.116 $\pm$ 0.013   &   0.069 $\pm$ 0.011\\
  -21.18   &   0.141 $\pm$ 0.025   &   0.167 $\pm$ 0.029   &   0.104 $\pm$ 0.021\\
  -20.80   &   0.186 $\pm$ 0.040   &   0.230 $\pm$ 0.045   &   0.135 $\pm$ 0.033\\
  -20.41   &   0.230 $\pm$ 0.047   &   0.289 $\pm$ 0.055   &   0.164 $\pm$ 0.036\\
  -20.01   &   0.264 $\pm$ 0.052   &   0.340 $\pm$ 0.062   &   0.185 $\pm$ 0.039\\
  -19.61   &   0.306 $\pm$ 0.050   &   0.400 $\pm$ 0.060   &   0.211 $\pm$ 0.037\\
  -19.22   &   0.336 $\pm$ 0.056   &   0.453 $\pm$ 0.067   &   0.233 $\pm$ 0.040\\
  -18.82   &   0.358 $\pm$ 0.062   &   0.505 $\pm$ 0.078   &   0.253 $\pm$ 0.042\\
  -18.41   &   0.372 $\pm$ 0.054   &   0.548 $\pm$ 0.069   &   0.269 $\pm$ 0.033\\
  -18.01   &   0.382 $\pm$ 0.057   &   0.584 $\pm$ 0.080   &   0.275 $\pm$ 0.031\\
  -17.61   &   0.396 $\pm$ 0.064   &   0.624 $\pm$ 0.083   &   0.280 $\pm$ 0.035\\
  -17.21   &   0.400 $\pm$ 0.070   &   0.650 $\pm$ 0.086   &   0.288 $\pm$ 0.044\\
  \cline{1-4}\\
  11.28   &   0.061 $\pm$ 0.009   &   0.062 $\pm$ 0.008   &   0.049 $\pm$ 0.026\\
  11.08   &   0.108 $\pm$ 0.016   &   0.112 $\pm$ 0.015   &   0.084 $\pm$ 0.022\\
  10.89   &   0.155 $\pm$ 0.029   &   0.164 $\pm$ 0.030   &   0.123 $\pm$ 0.022\\
  10.70   &   0.197 $\pm$ 0.039   &   0.226 $\pm$ 0.042   &   0.146 $\pm$ 0.031\\
  10.50   &   0.233 $\pm$ 0.047   &   0.287 $\pm$ 0.051   &   0.166 $\pm$ 0.041\\
  10.30   &   0.260 $\pm$ 0.049   &   0.343 $\pm$ 0.051   &   0.185 $\pm$ 0.045\\
  10.10   &   0.282 $\pm$ 0.052   &   0.401 $\pm$ 0.052   &   0.199 $\pm$ 0.046\\
  9.91   &   0.300 $\pm$ 0.057   &   0.461 $\pm$ 0.064   &   0.220 $\pm$ 0.047\\
  9.70   &   0.304 $\pm$ 0.057   &   0.507 $\pm$ 0.055   &   0.237 $\pm$ 0.050\\
  9.51   &   0.317 $\pm$ 0.054   &   0.557 $\pm$ 0.057   &   0.255 $\pm$ 0.046\\
  9.31   &   0.328 $\pm$ 0.054   &   0.587 $\pm$ 0.066   &   0.266 $\pm$ 0.041\\
  9.11   &   0.336 $\pm$ 0.053   &   0.610 $\pm$ 0.064   &   0.271 $\pm$ 0.039\\
  \enddata
  
  \tablecomments{Satellite fractions of galaxies  in the SDSS DR4, as function
    of galaxy  luminosity and stellar mass.   Results are listed  for all, red
    and  blue  galaxies, respectively.   Column  (1):  The average  luminosity
    ($\rmag$; upper  part) and  stellar mass ($\log  [M_*/h^{-2}\Msun]$; lower
    part) of galaxies. Columns (2-4):  the fractions of satellite galaxies for
    all,  red  and blue  galaxies  (averages  with  errors).  The  errors  are
    estimated  from the variances  between Samples  II and  III, and  are much
    larger    than   the    errors   obtained    from   our    200   bootstrap
    samples.}\label{tab:fsat}
\end{deluxetable}

The  satellite fraction  as function  of luminosity,  $f_{\rm sat}(L)$,  is an
important   quantity  for   a   proper  interpretation   of  measurements   of
galaxy-galaxy lensing (e.g., Guzik \& Seljak 2002; Mandelbaum \etal 2006; Yang
\etal  2006) and  pairwise  velocity  dispersion of  galaxies  (e.g., Jing  \&
B\"orner 2004; Slosar, Seljak \& Tasitsiomi 2006; van den Bosch et al.  2007a;
Li et al. 2007).  In addition, since the halo bias depends on halo mass (Mo \&
White 1996), and since a satellite  galaxy resides in a more massive halo than
a central galaxy of the same luminosity (van den Bosch \etal 2007b), the large
scale clustering  of galaxies of a  given luminosity also  depends strongly on
the  fraction  of  satellite  galaxies.   Here we  estimate  $f_{\rm  sat}(L)$
directly from  our group catalogue.  According  to the test we  carried out in
the previous section, this fraction can be determined relatively accurately.

In the left-hand  panel of Fig.~\ref{fig:f_sat} we show  $f_{\rm sat}(L)$ as a
function  of galaxy  luminosity. The  results are  plotted separately  for all
(solid  lines), red  (dashed lines)  and blue  (dotted lines)  galaxies. Since
fiber  collisions are  expected to  significantly impact  the number  of close
pairs, it can affect the satellite fractions $f_{\rm sat} (L)$.  To assess the
uncertainty induced by the fiber  collisions, we show results for both Samples
II  and III, as  thick and  thin lines,  respectively.  As  mentioned earlier,
Samples II  and III  may respectively under-  and over-estimate the  number of
group members because of their different treatments of fiber collisions.  Most
likely the  satellite fractions  of these two  extreme cases bracket  the true
satellite fractions.   As one can  see, the satellite fraction  decreases with
increasing luminosity,  from $\sim  40\%$ at $\rmag=-17.0$  to $\sim 5\%  $ at
$\rmag=-22$.  The satellite  fraction of red galaxies at  the faint end, $\sim
70\%$ at  $\rmag=-17.0$, is significantly  higher than that of  blue galaxies,
$\sim 30\%$ at $\rmag=-17.0$.  The  satellite fraction as a function of galaxy
stellar mass is shown in the right panel of Fig.~\ref{fig:f_sat}.  The overall
behavior here is  similar to that shown in the left  panel.  For reference, we
list  in Table  \ref{tab:fsat} the  fraction of  satellites as  a  function of
galaxy luminosity or stellar mass,  separately for all, red and blue galaxies.
The averages listed in this table  are the averages between Sample II and III,
while  the errors  quoted represent  the deviations  of the  samples  from the
average.

In  recent  years,  the  fraction  of  satellite  galaxies  has  been  studied
extensively using  HOD/CLF models  (Cooray 2006; Tinker  et al. 2007;  van den
Bosch et al.  2007a), and analyses of galaxy-galaxy lensing measurements (e.g.
Mandelbaum et al.  2006).  For comparison, we overplot in Fig.~\ref{fig:f_sat}
the  results obtained  by Mandelbaum  et al.   for early-type  galaxies (solid
circles) and  late-type galaxies (open triangles),  with error-bars indicating
the  95\% confidence level.   Although their  samples are  defined differently
from  ours (they  separate galaxies  into early-  and late-types  according to
galaxy  morphologies, while  we separate  galaxies according  to  colour), our
results match theirs quite well.

\section{Summary}
\label{sec_summary}

Using a large galaxy group catalogue  constructed from the SDSS Data Release 4
(DR4)  by Y07,  we have  investigated  various halo  occupation statistics  of
galaxies. In particular,  we have split the galaxy population  in red and blue
galaxies,  and in  centrals  and satellites,  and  determined the  conditional
luminosity  functions of  these  varies subsamples.   We  have also  presented
luminosity gap  statistics, satellite  fractions, and halo  occupation numbers
for  the galaxies  in our  group sample.  The main  results are  summarized as
follows:

\begin{enumerate}
  
\item The conditional luminosity  functions for central and satellite galaxies
can be well  modelled with a log-normal distribution  and a modified Schechter
form, respectively.   The corresponding best fitting parameters  are listed in
Table 1.
  
\item The average scatter of the log-normal luminosity distribution of central
galaxies decreases from $\sim 0.15$ dex at the massive end ($\log [M_h/\msunh]
\ga 13.5$)  to $\sim  0.1$ dex at  the low  mass end ($\log  [M_h/\msunh] \sim
12.0$). However, due  to the method used to assign halo  masses to the groups,
at the low mass end this should be considered a lower limit on the true amount
of scatter.
  
\item  The slope of  the relation  between the  average luminosity  of central
galaxies (in the  $^{0.1}r$-band) and halo mass, ${\rm  d}\log L_c/{\rm d}\log
M_h$, decreases  from $\sim  0.68$ for $\log  [M_h/\msunh] \ll 12.5$  to $\sim
0.17$ for$\log [M_h/\msunh]  \gg 12.5$. For the stellar  masses of the central
galaxies we obtain that ${\rm  d}\log M_{*,c}/{\rm d}\log M_h$, decreases from
$\sim  1.83$  for  $\log  [M_h/\msunh]  \ll  12.1$  to  $\sim  0.22$  for$\log
[M_h/\msunh] \gg 12.1$.
  
\item  The halo (group)  occupation numbers  of satellite  galaxies accurately
follow Poisson  statistics. Since the  same applies to dark  matter sub-halos,
this supports the standard picture that satellite galaxies are associated with
dark matter sub-halos.
  
\item In massive halos with  masses $M_h\ga 10^{14}\msunh$ roughly 85\% (80\%)
of the central  (satellite) galaxies are red. These  red fractions decrease to
50\% (40\%) in halos with masses $M_h \sim 10^{12}\msunh$.
  
\item By comparing  the scatter in the luminosities of  BCGs to the luminosity
difference between the BCG and its  brightest satellite, we find that the BCGs
form a  `special' subclass, in that  their luminosities can  not be considered
the extreme  values of the distribution of  satellite luminosities, expecially
in halos with masses $M_h \la 10^{14.0} \msunh$.
  
\item  The fractions  of fossil  groups, which  are defined  as groups  a with
luminosity gap  $\log L_1 -  \log L_2 \ge  0.8$, decreases with  increasing of
halo  mass from  18\%-60\% in  halos with  $M_h \sim  10^{13}\msunh$  to $\sim
2.5\%$ in halos with $M_h \sim 10^{14}\msunh$.
 
\item The satellite  fractions obtained from our group  catalogue as functions
of both  luminosity and stellar  mass (listed in Table~\ref{tab:fsat})  are in
good agreement with independent constraints from analyses of galaxy clustering
and galaxy-galaxy lensing.

\end{enumerate}

These results can be used  to constrain the various physical processes related
to galaxy formation  and to interpret the various  statistics used to describe
large scale structures (e.g.,  galaxy correlation functions, pairwise velocity
dispersions, etc.).  Most of our  findings are in good agreement with previous
studies (e.g.  Y05c,  Zandivarez et al.  2006; Robotham et  al.  2006) and can
be linked to the semi-analytical modelling of galaxy formations (e.g., Kang et
al. 2005; Zheng et al.  2005; Bower et al. 2006; Croton et al.  2006; De Lucia
et al. 2007).   The luminosity and stellar  mass gap can be used  to probe the
specific  formation properties  of central  galaxies (e.g.,  Vale  \& Ostriker
2007). The fraction of the red  and blue populations for central and satellite
galaxies can be  used to probe the color evolution  of satellite galaxies (ven
den Bosch et al.  2007b).


\section*{Acknowledgments}

We thank J.P.  Ostriker and Zheng Zheng for helpful comments, and Cheng Li for
the galaxy  color bi-normal fitting. XY  is supported by the  {\it One Hundred
Talents} project, Shanghai Pujiang  Program (No.  07pj14102), 973 Program (No.
2007CB815402), the CAS Knowledge  Innovation Program (Grant No.  KJCX2-YW-T05)
and grants from NSFC (Nos.  10533030, 10673023). HJM would like to acknowledge
the support of NSF AST-0607535,  NASA AISR-126270 and NSF IIS-0611948. Funding
for the SDSS and SDSS-II has been provided by the Alfred P.  Sloan Foundation,
the  Participating Institutions,  the  National Science  Foundation, the  U.S.
Department of  Energy, the National Aeronautics and  Space Administration, the
Japanese  Monbukagakusho, the  Max Planck  Society, and  the  Higher Education
Funding Council for England.  The  SDSS Web Site is http://www.sdss.org/.  The
SDSS is managed by the Astrophysical Research Consortium for the Participating
Institutions.   The  Participating Institutions  are  the  American Museum  of
Natural  History,  Astrophysical   Institute  Potsdam,  University  of  Basel,
Cambridge University, Case Western  Reserve University, University of Chicago,
Drexel  University, Fermilab,  the  Institute for  Advanced  Study, the  Japan
Participation Group, Johns Hopkins University, the Joint Institute for Nuclear
Astrophysics, the Kavli Institute for Particle Astrophysics and Cosmology, the
Korean Scientist Group,  the Chinese Academy of Sciences  (LAMOST), Los Alamos
National  Laboratory,  the  Max-Planck-Institute  for  Astronomy  (MPIA),  the
Max-Planck-Institute for Astrophysics (MPA), New Mexico State University, Ohio
State  University,   University  of  Pittsburgh,   University  of  Portsmouth,
Princeton University, the United  States Naval Observatory, and the University
of Washington.


\label{lastpage}

\end{document}